\documentclass[a4paper,12pt]{llncs}
\usepackage[utf8]{inputenc}
\usepackage[T1]{fontenc}
\usepackage{amsmath}
\usepackage{amsfonts}
\usepackage{amssymb}
\usepackage{authblk}
\usepackage{fullpage}

\usepackage{multicol}

\usepackage{ifpdf}
\usepackage{url}

\ifpdf
 \usepackage[pdftex]{graphicx}
 \usepackage[pdftex]{color}
\else
 \usepackage{graphicx}
 \usepackage{color}
\fi
\usepackage{wrapfig}
\usepackage{verbatim}
\usepackage{tikz-cd}

\usepackage{tcolorbox}
\definecolor{aliceblue}{rgb}{0.94, 0.97, 1.0}

\renewcommand\refname{}

\newcommand{\rem}[1]{}

\usepackage{natbib}
	\setcitestyle{authoryear,round,aysep={}}

\begin{document}

\thispagestyle{empty}


{\Large \bf The Coconut Model with Heterogeneous Strategies and Learning}\vspace{6pt}\\
{\large S. Banisch\footnote{Correspondence should be send to \email{sven.banisch@UniVerseCity.de}}
, E. Olbrich} \vspace{2pt}\\
Max Planck Institute for Mathematics in the Sciences, Leipzig, Germany
\begin{center}
\vspace{-18pt}
\line(1,0) {450}
\end{center}




\begin{abstract}
In this paper, we develop an agent-based version of the Diamond search equilibrium  model -- also called Coconut Model.
In this model, agents are faced with production decisions that have to be evaluated based on their expectations about the future utility of the produced entity which in turn depends on the global production level via a trading mechanism.
While the original dynamical systems formulation assumes an infinite number of homogeneously adapting agents obeying strong rationality conditions, the agent-based setting allows to discuss the effects of heterogeneous and adaptive expectations and enables the analysis of non-equilibrium trajectories.
Starting from a baseline implementation that matches the asymptotic behavior of the original model, we show how agent heterogeneity can be accounted for in the aggregate dynamical equations.
We then show that when agents adapt their strategies by a simple temporal difference learning scheme, the system converges to one of the fixed points of the original system.
Systematic simulations reveal that this is the only stable equilibrium solution.
\end{abstract}

\begin{center}
\line(1,0) {450}
\end{center}
\vspace{-8pt}
The results of this paper have been subsequently presented at the 2015 and 2016 conference on Artificial Economics (\email{http://www.artificial-economics.org}).
We acknowledge the valuable feedback from the participants of these two events.
The work was supported from the European Community's Seventh Framework Programme (FP7/2007-2013) under grant agreement no.~318723 (MatheMACS). S.B. also acknowledges financial support by the Klaus Tschira Foundation.
\vspace{-10pt}
\begin{center}
\line(1,0) {450}
\end{center}





\section{Introduction}

Aggregate dynamical equations derived from individual-based models are usually exact only if the micro-level dynamics satisfy certain rather restrictive symmetry conditions brought about by homogeneity of the agent properties and interaction rules \citep{Banisch2015springer}.
Micro-level heterogeneity or complex interaction structures often lead to correlations in the dynamics that a macro-level description may not account for.
That is, it is not possible to exactly describe the system evolution as a closed set of macroscopic equations.
For this reason, it is important to understand the consequences of heterogeneous agent properties on macroscopic formulations of the system dynamics.
In this paper, we present results into that direction for the Diamond search equilibrium model \citep{Diamond1982,Diamond1989} -- also known as Coconut Model -- which has been introduced in 1982 by the 2010 Nobel laureate Peter Diamond as a model for an economy with trade frictions.

Imagine an island with $N$ agents that like to eat coconuts.
They search for palm trees and harvest a nut from it if the tree is not too tall, meaning that its height does not exceed an individual threshold cost ($c_{tree} < c_i$).
However, in order to consume the nut and derive utility $y$ from this consumption agents have to find a trading partner, that is, another agent with a nut.
Therefore, the agents have to base their harvest decision \emph{now} (by setting $c_i$) on their expectation to find a trading partner \emph{in the future}.
Or, less metaphorically, agents are faced with production decisions that have to be evaluated based on their expectations about the future utility of the produced entity which in turn depends on the global production level via a trading mechanism.
For this reason, the Coconut Model is useful not only for the incorporation of heterogeneity, but also for the analysis of adaptive agents that -- rationally or not -- have to form expectations about the future system state in order to evaluate their decision options. 

In the original papers \citep{Diamond1982,Diamond1989} this problem of inter-temporal optimization is formulated using dynamic programming principles and the Bellmann equation in particular. 
The author(s) arrive at a differential equation (DE) that describes the evolution of the cost threshold along an optimality path (where the individual thresholds are all equal $c_i = c$) which is coupled to a second DE describing the evolution of the number of coconuts in the population.
However, knowing the optimal dynamics, that is, the differential equations that an optimal solution has to fulfill, is not sufficient to study problems such as equilibrium selection or stability in general, because the optimality conditions do not say anything about the behavior of the system when it is perturbed into a suboptimal state. 
On the other hand, the Bellmann equation is also at the root of reinforcement learning algorithms and temporal difference (TD) learning in particular which are known to converge to this optimality under certain conditions \citep{Sutton1998}.
The incorporation of learning in the agent-based version of the Coconut Model and the assessment of its adequacy by comparison to the original solution is a second contribution of this paper.

The necessity to take into account not only the result of rational choice but to focus more on the processes that may lead to it has been pointed at by Simon almost 40 years ago \citep{Simon1978}.
Around ten years later, the notion of artificial adaptive agents has been proposed by 
\cite{Holland1991} who define an \emph{adaptive} agent by two criteria: (1.) the agent assigns a value (fitness, accumulated reward, etc.) to its actions, and (2.) the agent intends to increase this value over time (p. 365).
Virtually all models with adaptive agents proposed since then follow these principles.
In genetic algorithms, for instance, an evolutionary mechanism is implemented by which least fit strategies are replaced by fitter ones and genetic operators like recombination and mutation are used to ensure that potential improvements are found even in high-dimensional strategy spaces, \citep[e.g.][]{Holland1975,Vriend2000}.
Another approach which became prominent during the last years could be referred to as strategy switching \citep[e.g.][]{Brock1997,Hommes2013,Landini2015}.
Here agents constantly evaluate a set of predefined decision heuristics by reinforcement mechanisms and chose the rule that performs best under the current conditions.
The reader may be referred to the excellent introductory chapter to >>Behavioral Rationality and Heterogeneous Expectations in Complex Economic Systems<< by \cite{Hommes2013} for a very instructive account.

The TD approach used here differs mildly from these models but fits well with the abstract specification of adaptive behavior proposed in \cite{Holland1991}.
In our case agents learn the value associated to having and not having a coconut in form of the expected future reward and use these values to determine their cost threshold $c_i$.
That is, agents are forward-looking by trying to anticipate their potential future gains.
While checking genetic algorithms or strategy switching methods in the context of the Coconut Model is an interesting issue for future research, in this first paper we would like to derive an agent-based version of the model that is as closely related to the original model as possible.
The motivation behind this is well-captured by a quote from \cite{Holland1991} (p. 366):
\begin{quote}
>>As a minimal requirement, wherever the new approach overlaps classical theory, it must include verified results of that theory in a way reminiscent of the way in which the formalism of general relativity includes the powerful results of classical physics.<<
\end{quote}

To our opinion, this relates to the tradition in ABM research to verify and validate computational models by replication experiments \citep[e.g.][]{Axtell1996,Hales2003,Grimm2006,Wilenski2007}. 
The main idea is that the same conceptual model implemented on different machines and possibly using different programming languages should always yield the same behavior.
To our point of view this attempt to develop scientific standards for ABMs should not be restricted to comparing different computer implementations of the same conceptual model, but it should also aim at aligning or comparing the results of an ABM implementation to analytical formulations of the same processes.
At least, whenever such descriptions are available.

For the Coconut Model this is the case, and for its rich and sometimes intricate behavior on the one hand and the availability of a set of analytical results on the other the model is well--suited as a testbed for this kind of model comparison.
In particular when it comes to extend a model that is formulated for an idealized homogeneous population so to incorporate heterogeneity at the agent level, we should make sure that it matches with the theoretical results that are obtained for the idealized case.

Hence, the main objective of this paper is to derive an agent-based version of the Coconut Model as conceived in the original papers \citep{Diamond1982,Diamond1989} where the model dynamics have been derived for an idealized infinite and homogeneous population.
We will see (in Section \ref{sec:homogeneous}) that this implementation does not lead to the fixed point(s) of the original system.
In order to align the ABM to yield the right fixed point behavior we have at least two different options which again lead to slight differences in the dynamical behavior that are not obvious from the DE description.

This then allows us to study the effects that result from deviating from the idealized setting of homogeneous strategies, which we address in Section \ref{sec:heterogeneous}.
Heterogeneity may arise from the interaction structure, the information available to the agents as well as from heterogeneous agent strategies as an effect of learning in finite populations, and we will concentrate on the latter here.
In particular, we will show that heterogeneity can by accounted for  in a macroscopic model formulation by the correction term introduced in \cite{Olbrich2012}.

A similar program shall be followed in Section \ref{sec:learning} where agents use TD learning to learn the optimal strategy.
As the learning scheme used in this paper can in fact be derived from the Bellman equation used to set up the original model, an agent population that adapts according to this method should converge to the same equilibrium solution in a procedural way.
However, as such an approach implements rationality as a process \citep{Simon1978} it describes the route to optimality and allows to analyze questions related to equilibrium selection and stability.
We shall now describe the original model more carefully.

\section{Description of the Original Model}

Consider an island populated by $N$ agents. 
On the island there are many palm trees and the agents wish to consume the coconuts that grow on these trees.
The probability that agents find a coco tree is denoted by $f$ and harvesting a nut from the tree bears a cost $c_{tree}$ (the metaphor is the height of the tree) that is described by a cumulative distribution $G(c)$ defining the probability that the cost of a tree $c_{tree} < c$.
In what follows we consider that the costs that come with a tree are uniformly distributed in the interval $[c_{min},c_{max}]$ such that $G(c) = (c-c_{min})/(c_{max} - c_{min})$.
Agents cannot store coconuts such that only agents without a nut may climb and harvest a new one.
On encountering a tree (with probability $f$), an agent without coconut climbs if the cost $c_{tree}$ of harvesting the tree they encountered does not exceed a strategic cost threshold $c_i$ (referred to as strategy) above which the agent assumes that harvesting the coconut would not be profitable.
In other words, the probability with which an agent without nut will harvest a tree is given by $f G(c_i)$.
In the original model, the agent strategy is endogenously determined (and then written as $c_i^t$) as described below.

In what follows, we denote the state of an agent $i$ by the tuple $(s_i,c_i)$ where $s_i \in \{0,1\}$ encodes whether agent $i$ holds a coconut or not and $c_i$ is the strategy of the agent.
We define the macroscopic quantities $e = \sum\limits_{i}^{} s_i$ and $\epsilon = e/N$ with the first being the number of coconuts in the population and the second the ratio of agents having a coconut, respectively. 
The time evolution in the limit of an infinite population can then be written as a differential equation (DE)
\begin{equation}
\dot{\epsilon} = f (1-\epsilon) G(c^*) - \epsilon^2 
\label{eq:evoDGL}
\end{equation}
where the first term corresponds to the rate at which agents harvest a coconut and the second to trading with $\epsilon^2$ being the probability of randomly choosing two agents with coconut \cite{Diamond1982}.
$c^*$ corresponds to the optimal strategy of the agents. It is assumed to be homogeneous for the population but endogenously defined (time-dependent) in the original works of Diamond and co-workers. $G(c^*)$ is the fraction of trees that will be harvested at this cost threshold.

A crucial ingredient of the coconut model is that agents are not allowed to directly consume the coconut they harvested.
They rather have to search for a trading partner, that is, for another agent that also has a coconut.
The idea behind this is that agents have to find buyers for the goods they produce.
If an agent that possesses a nut encounters another agent with a nut both of them are supposed to consume instantaneously and derive each a reward of $y$ from this consumption.
In effect, this means that the expected value of climbing a tree depends on the total number of coconuts in the population or, more precisely, on the time agents have to wait until a trading partner will be found.

Rational agents are assumed to maximize their expected future utility 
\begin{equation}
\nonumber
V_i(t) = \mathbb{E} \int\limits_{t}^{\infty} e^{-\gamma (\tau - t)} r_i(\tau) d\tau
\end{equation}
where $r_i(\tau)$ corresponds to the cost of climbing or respectively to the utility $y$ from consumption of agent $i$ at time $\tau$ and $\gamma$ to the discount factor.
A fully rational agent has to find the strategy $c^*_i$ that maximizes its expected future reward and, since agents cannot consume their coconut instantaneously, this reward depends on his expectation about their trading chances.
This can be formulated as a dynamic programming problem with $dV_i(t)/dt = - \langle r_i(t)\rangle + \gamma V_i(t)$.
Considering that there are two states (namely, $s_i = 0$ or $s_i = 1$) there is a value associated to having ($V_i(s_i=1,t) := V_i^t(1)$) and to not having  ($V_i(s_i=0,t) :=V_i^t(0)$)  a coconut at time $t$.
As a rational agent accepts any opportunity that increases expected utility, a necessary condition for an optimal strategy is $c_i^* = V_i^t(1) - V_i^t(0)$.
By this reasoning, assuming homogeneous strategies $c_i^*=c^*$ Diamond derives another DE that describes the evolution of the optimal strategy
\begin{equation}
\frac{dc^*}{dt} = \gamma c^* + \epsilon(c^* - y) +f \left[ \int_{0}^{c^*} (c^* - c) dG(c) \right]
\label{eq:evoDGLStrat}
\end{equation}

The model (\ref{eq:evoDGL})-(\ref{eq:evoDGLStrat}) is interesting due to its a rather rich solution structure and because many macroeconomic models incorporating inter-temporal optimization have a similar form.
First, the model may give rise to a market failure equilibrium which arises as a self-fulfilling prophecy where everyone believes that nobody else invests (climbs the coco trees).
It has multiple equilibria and it has been shown in \cite{Diamond1989} that the model could also exhibit cyclic equilibria.
This has been interpreted as an abstract model for endogenous business cycles.
A complete stability analysis of the system (\ref{eq:evoDGL}) and (\ref{eq:evoDGLStrat}) has been presented by \cite{Lux1992}.

The formulation of the model as a two-dimensional system of DEs (\ref{eq:evoDGL})-(\ref{eq:evoDGLStrat}) \citep{Diamond1982,Diamond1989,Lux1992} assumes an infinite and homogeneously adapting population.
While many aspects of the DE system have been understood rather well using the corresponding analytical tools, some other important aspects of general theoretical interest cannot be addressed within this setting.
This includes equilibrium selection, but also stability and out-of-equilibrium perturbations, bounded or procedural rationality \citep{Simon1978} and learning, and finally the influence of micro-level heterogeneity.
In order to address the first of these points -- equilibrium selection -- a reexamination of the model with finite populations had been undertaken in \cite{Aoki2000}.
The authors relate equilibrium selection to the finite-size stochastic fluctuations and show that transitions between different basins of attraction take place with positive probability. 
The benefit of implementing Diamond's model as an ABM is that we can eventually relax the other assumptions as well.

\section{Homogeneous Strategies}
\label{sec:homogeneous}

As argued in the introduction, when reimplementing the model as an ABM, we should make sure that the model incorporates the solution(s) of the original model.
In other words, that the ABM with the original idealizations maintained leads to the same behavior.
Therefore, let us first look at the dynamics of the model when all agents adopt the same strategy $c_i = c$. 
The fixed point solution of the DE (\ref{eq:evoDGL}) is
\begin{equation}
\epsilon^*(c) = \frac{f G(c)}{2} \left[ \sqrt{1 + \frac{4}{f G(c)}} - 1 \right]
\label{eq:fix}
\end{equation}
which characterizes the equilibrium point $\epsilon^*$ for a given $c$.
If we implement the coconut model  with homogeneous strategies $c$ in form of an ABM we can expect that the average level of coconuts in the population approaches and fluctuates around the point $\epsilon^*(c)$.

\subsection{Model Alignment}

\subsubsection{Intuitive Implementation}

Intuitively the Diamond model as set up in the original paper \cite{Diamond1982} could be implemented in the following way (we refer to this >>intuitive<< implementation as \verb|IM|):\vspace{4pt}\\
\begin{tcolorbox}[width=\textwidth,colback={aliceblue}]  
\begin{itemize}
\item[(0)] 
Random set-up: set strategies $c^0_i = c_i$ and initial states $s^0_i$ according to the desired initial distribution.
\item[(1)] Iteration loop:
	\begin{itemize}
	\item[(a)] random choice of an agent $i$ with probability $\omega(i) = 1/N$
	\item[(b)] \verb|if| $s^t_i = 0$ climb a coco tree with probability $f G(c_i)$ and harvest a nut, i.e., $s_i^{t+1} = 1$
	\item[(c)] \verb|else if| $s_i^t = 1$
		\begin{itemize}
		\item choose a second agent $j$ with probability $\omega(j) = 1/(N-1)$ (random matching)
		\item \verb|if| $s_j^t = 1$ trade such that $s_i^{t+1} = 0$ and $s_j^{t+1} = 0$
		\end{itemize}
	\end{itemize}
\end{itemize}
\end{tcolorbox}
Some short comments on this procedure are in order.
For the initialization step (0) we usually define a desired level of coconuts in the initial population ($\epsilon^0$) and a random initial state adhering to this level is obtained by letting each agent have a coconut with probability $\epsilon^0$.
The initialization of the strategies $c_i$ will be according to the different scenarios described in the next section but in general the strategies will ly in the interval $[c_{min},c_{max}]$.
In this section, the strategies are homogeneous and do not change over time so that all agents have identical strategies during all times.
Point (1a) in the iteration process means that at each step we randomly choose one agent from the population.
Notice that this means that within $N$ iteration steps some agents may be chosen more than once whereas others might not be chosen at all.
For point (1b) the climbing decision with probability $fG(c_i)$ is evaluated by drawing two random numbers, one for the rate of coco trees $f$ and another for the cost of the coconut which is uniformly distributed in the interval $c_{tree} \in [c_{min},c_{max}]$ for all the experiments we perform throughout this paper.
Agents will climb the tree if $c_i \leq c_{tree}$.
Finally, if agent $i$ has a nut to trade (1c), a second agent $j$ is randomly chosen from the remaining agent set which means that there are no specific trading relations between the agents that direct the search for trading partners. 

\begin{figure}[h]
\centering
\includegraphics[width=0.69\linewidth]{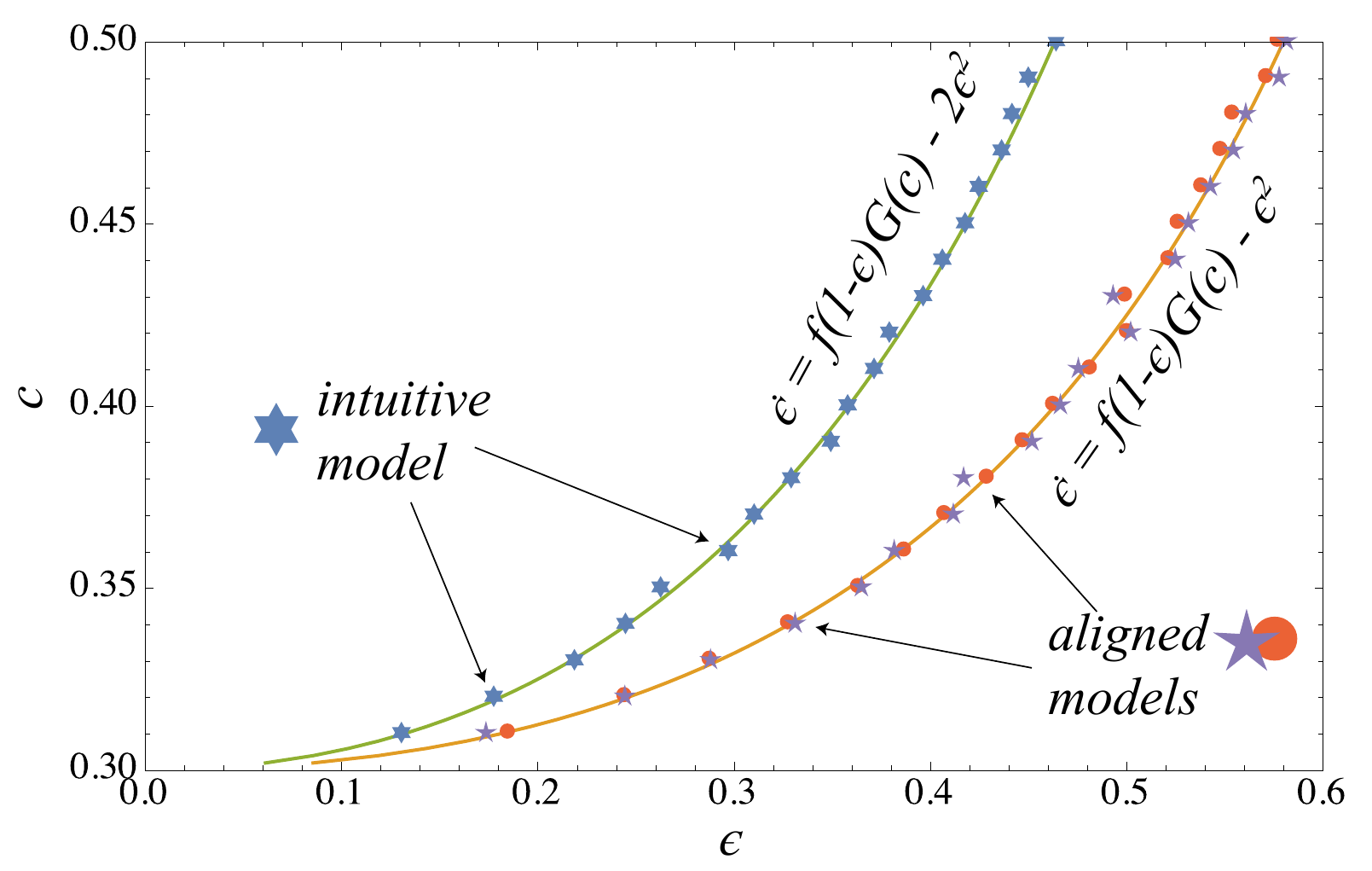}  
\caption{Comparison of the mean level of coconuts (in the stationary regime) of the intuitive implementation of the Diamond model with the original differential equation and an adapted equation that accounts for the fact that two agents may trade but only one agent may climb in one time step. The behavior of two ABM implementations aligned to the original DE is also shown.}
\label{fig:ComparisonIM}
\end{figure}

In this section, we compare the outcome of the simulation model as specified above with homogeneous strategies $c_i = c$ to the behavior predicted by the DE (\ref{eq:evoDGL}).
We look at a small system of only 100 agents in order to see how well the theoretical results obtained for infinite populations approximates the small-scale system.
For the comparison the system is initialized with $\epsilon^0 = 0$ (no coconuts in the population) and run for 4000 steps to reach the stationary regime.
Then another 10000 update steps are performed during which we measure the mean level of coconuts $\epsilon$ in the population.
We run a series of such experiments for different strategy values $c$ in between $c_{min}$ and $c_{max}$.
The results of this are shown by the stars points in Fig. \ref{fig:ComparisonIM}.

In addition, we also show the strategy dependent solution curve $\epsilon^*(c)$ (\ref{eq:fix}) obtained for the original DE (orange curve).
It becomes clear that for the intuitive implementation the average coconut level is considerably below the DE solution.
This is due to the fact that in a single time step, on average, two agents may trade but only one agent may climb a coco tree and harvest a nut which means that the rate $-\epsilon^2$ actually underestimates the decrease of coconuts by a factor of two.
Consequently, we should correct this term for $-2 \epsilon^2$ in order to obtain a DE corresponding to the behavior of the ABM.
That is,
\begin{equation}
\dot{\epsilon} = f (1-\epsilon) G(c) - 2\epsilon^2 
\label{eq:evoDGL2}
\end{equation}
yielding the fixed point curve
\begin{equation}
\epsilon^*(c) = \frac{f G(c)}{4} \left[ \sqrt{1 + \frac{8}{f G(c)}} - 1 \right]
\label{eq:fix2}
\end{equation}
which is also shown (green curve) in Fig. \ref{fig:ComparisonIM}.
It becomes clear that the simulation of the \verb|IM| with homogeneous strategies reproduces fairly well the fixed point solution (\ref{eq:fix2}) of the adjusted DE.

Notice that the adjusted DE (\ref{eq:evoDGL2}) does not change the behavior of the system (\ref{eq:evoDGL})-(\ref{eq:evoDGLStrat}) in a qualitative way.
Notice also that instead of multiplying $-\epsilon^2$ by two, we could also rescale $f$ to $f/2$ so that the dynamical structure of the model is not really affected.
For the experiments in Section \ref{sec:heterogeneous} we will stick to the intuitive model implementation (\verb|IM|) as described above and use (\ref{eq:fix2}) for the comparison with the DE description.

\subsubsection{Two Ways of Aligning the ABM}
\label{sec:aligning}

In order to obtain an ABM that matches well with the original DE description, we have at least two different options.
On the one hand, we could allow two agents to climb at each time step by changing the iteration scheme to\vspace{6pt}\\
\begin{tcolorbox}[width=\textwidth,colback={aliceblue}]  
\begin{itemize}
\item[(1)] Iteration loop:
	\begin{itemize}
	\item[(a)] random choice of an agent pair $(i,j)$ with probability $\omega(i,j) = 1/N(N-1)$
	\item[(b$_1$)] \verb|if| $s_i^t = 0$ with probability $f G(c_i)$ climb tree and harvest, i.e., $s_i^{t+1} = 1$
	\item[(b$_2$)] \verb|if| $s_j^t = 0$  with probability $f G(c_j)$ climb tree and harvest, i.e., $s_j^{t+1} = 1$
	\item[(c)] \verb|else if| $s_i^t s_j^t = 1$ trade such that $s_i^{t+1} = 0$ and $s_j^{t+1} = 0$
	\end{itemize}
\end{itemize}
\end{tcolorbox}
Let's refer to this scheme as aligned model one (\verb|AM1|).

On the other hand, we could decide that an agent $i$ with a nut ($s^t_i=1$) trades that nut with a probability proportional to the number $\epsilon^t$ of coconuts in the population so that at most one coconut will be cleared at a single time step.\footnote{A decision of that kind seems to be made in \cite{Diamond1982}, see footnote 4.}
The iteration (we shall call this scheme \verb|AM2|) becomes:\vspace{6pt}\\
\begin{tcolorbox}[width=\textwidth,colback={aliceblue}]  
\begin{itemize}
\item[(1)] Iteration loop:
	\begin{itemize}
	\item[(a)] random choice of an agent $i$ with probability $\omega(i) = 1/N$
	\item[(b)] \verb|if| $s_i^t = 0$ with probability $f G(c_i)$ climb tree and harvest, i.e., $s_i^{t+1} = 1$
	\item[(c)] \verb|else| trade (consume) with probability $\epsilon^t$ such that $s_i^{t+1} = 0$
	\end{itemize}
\end{itemize}
\end{tcolorbox} 
In a sense, this mechanism does not really comply with what we usually do in an ABM.
However, it is a way to reproduce the fixed point solution (\ref{eq:fix}) of the original DE, which is shown in Fig. \ref{fig:ComparisonIM} as well.
Indeed, the two aligned schemes match very well with the DE solution.
\verb|AM2| will be used in Section \ref{sec:learning} where we incorporate learning dynamics into the model.

\subsection{Finite-Size Markov Chain Formulation}

In addition to the DE and the ABM formulation we shall consider here a third description in form of a Markov chain (MC).
For homogeneous populations one can show that an MC description accounting for the transitions between the different numbers of coconuts in the population provides an exact formulation of a finite size ABM with $N$ agents, \cite{Banisch2015springer}.
In the finite size case, the possible $\epsilon$ are given as the finite set $\{ 0/N, 1/N,\ldots, N/N \}$ which we shall write as $\mathcal{Y} = \{0,1,\ldots,e,\ldots,N\}$ to simplify notation. We denote by $e$ the (absolute) number of agents with a coconut (i.e., $e = \sum_i s_i$).
For the intuitive scheme, the transition probabilities for the \verb|IM| are given by
\begin{eqnarray}
P(e+1 | e) = f \frac{(N-e)}{N} G(c) \nonumber\\
P(e-2 | e) = \left(\frac{e (e-1)}{N(N-1)}\right)\nonumber\\
P(e | e) = 1- P(e+1 | e) - P(e-2 | e)
\label{eq:homoMC}
\end{eqnarray}
and zero elsewhere.
Notice that trading decreases the number of coconuts by two whereas only one coconut can be harvested in a single transition as implemented by the intuitive model.
If, instead, we choose $P(e-1 | e)=\left(\frac{e (e-1)}{N(N-1)}\right)$ and $P(e-2 | e) = 0$ we obtain a MC representation that is aligned to the \verb|AM2| scheme with only one agent consuming a nut on trade.

\begin{figure}[h]
\centering
\includegraphics[width=0.66\linewidth]{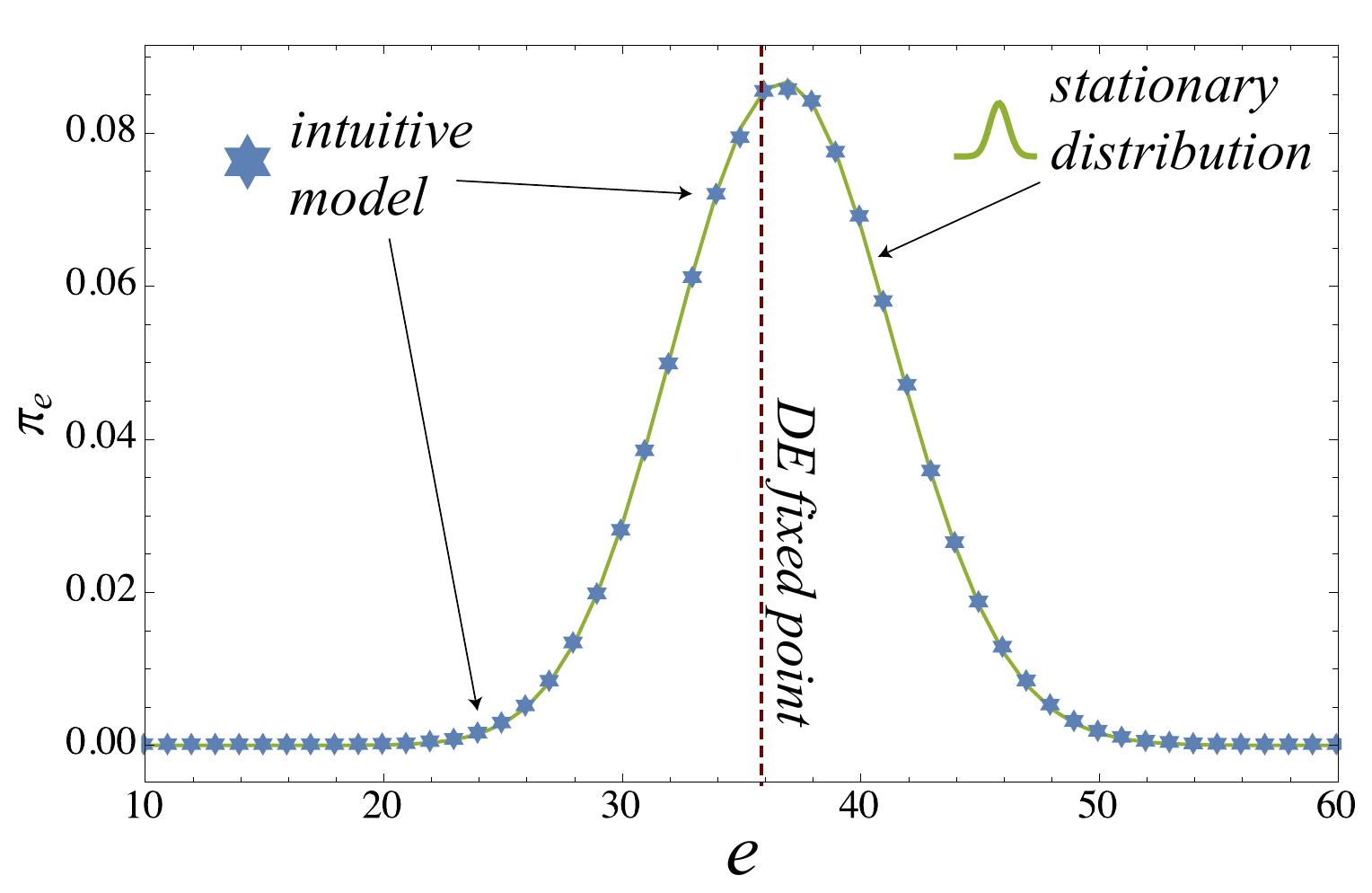} 
\caption{Stationary distribution of the MC for $N = 100$, $f = 0.8$ and $c=0.4$ compared to the frequencies observed in 10 realizations (each 100000 steps) of the simulation model.}
\label{fig:pi100}
\end{figure}

A MC representation can be useful because it provides us with a good understanding about the finite size fluctuations around the equilibrium point.
In Fig. \ref{fig:pi100}, this is shown by comparing the observed coconut level in the intuitive model to the stationary vector of (\ref{eq:homoMC}) for the homogeneous strategy $c = 0.4$.
Notice, however, that the peak in the MC stationary vector and the solution of the DE (\ref{eq:fix2}) do not match precisely.
This is due to the fact that the trading probabilities (with $\epsilon^2 = (e^2/N^2)$) in the mean-field formulation do not explicitly exclude self-trading, whereas this is excluded in the MC formulation as well as in the ABM (and therefore the respective probability is $(e(e-1)/N(N-1))$).
While the difference between the two probabilities is neglectable if $N$ is large, there is a notable effect for $N = 100$.

\section{Heterogeneous Strategies}
\label{sec:heterogeneous}

We now relax the assumption of homogeneous strategies by assigning a random individual strategy $c_i$ to each agent drawn from a certain probability distribution.
As before, the strategies are fixed for the entire simulation. 
The four different scenarios considered in the second part of this section differ with respect to the distribution from which the individual strategies are drawn (in general, in the interval $[c_{min},c_{max}]$).
Before looking at these specific scenarios, however, we derive a correction term that accounts for the effect of strategy heterogeneity.

\subsection{Heterogeneity Correction}
\label{sec:hetCorr}

Namely, if strategies are different we can expect that those agents in the population with a lower $c_i$ will also climb less often and are therefore less often with a coconut $s_i = 1$.
That is, there is a correlation between the agent strategy $c_i$ and the probability $Pr(s_i = 1)$ that an agent has a coconut.

To account for this we have to consider that the rate $P(e+1 | e)$ that an agent of the population will climb from one time step to the other is given by 
\begin{equation}
P(e+1 | e) = \frac{f}{N} \sum\limits_{i=1}^{N} (1-s_i)G(c_i)
\end{equation}
The equation covering the homogeneous case (\ref{eq:homoMC}) is satisfied because $G(c_i) = G(c)$ is equal for all agents and can be taken out of the sum.
For heterogeneous strategies this is not possible but we can come to a similar expression by formulating $P(e+1 | e)$ as the expected value (denoted as $\langle \rangle$)
\begin{equation}
\begin{array}{ll}
P(e+1 | e) & = f \langle(1-s_i) G(c_i)\rangle \vspace{6pt}\\
 & = f \left(\langle G(c_i)\rangle - \langle s_i G(c_i)\rangle \right)\vspace{6pt}\\
 & = f \left[\langle G(c_i)\rangle - \langle s_i\rangle \langle G(c_i)\rangle  - \sigma[s_i,G(c_i)] \right]\vspace{6pt}\\
 & = f\left[ (1- \langle s_i\rangle) \langle G(c_i)\rangle - \sigma[s_i,G(c_i)] \right],
\end{array}
\end{equation}
where $\sigma[s_i,G(c_i)]$ is the covariance between agent states $s_i$ and individual climbing probability $G(c_i)$.
With $\langle G(c_i)\rangle = G(\bar{c})$ and $\langle s_i\rangle = e/N$ (the former being true due to the linearity of $G$ and the latter by definition) this yields 
\begin{equation}
P(e+1 | e)  
= f\left[ \underbrace{\frac{N-e}{N} G(\bar{c})}_{\text{mean term}} - \underbrace{\sigma(s_i,G(c_i))}_{\text{covariance correction term}} \right]
\label{eq:corrMC}
\end{equation}
which corresponds to the original term (\ref{eq:homoMC}) except for the additional covariance term $\sigma(s_i,G(c_i))$.
Notice that the covariance depends on the individual agent states ($s_i$ and $G(c_i)$) so that the transition rate is no longer a pure function of the macroscopic average state.

This correction term accounting for the correlation between the agent strategy and state should also be included into the infinite-size formulation (\ref{eq:evoDGL2}).
This is accomplished by \cite{Olbrich2012}
\begin{equation}
\dot{\epsilon} = f \left[ (1-\epsilon) G(\bar{c}) - \sigma(t)\right]- 2 \epsilon^2.
\label{eq:evoDGLKorrT}
\end{equation}
Note that covariance term $\sigma(t)$ in Eq. (\ref{eq:evoDGLKorrT}) is here written as a time-dependent parameter because it depends on the actual microscopic agent configuration at time $t$ to be computed.
In other words, this is no longer a closed description in the macroscopic variable $\epsilon$ as it is coupled to the evolution of the heterogeneity in the system contained in the covariance term.

\subsection{Time Evolution of the Heterogeneity Term}

One way to address this issue is to derive an additional differential equation that describes the time evolution of $\sigma$ and is coupled to (\ref{eq:evoDGLKorrT}).
For this purpose let us denote the probability for an agent $i$ to be in state $s_i = 1$ as $Pr(s_i = 1) := p_i$.
The evolution of this probability is described by
\begin{equation}
\dot{p}_i=f (1-p_i) G(c_i)-\epsilon p_i
\label{eq:evoPropI}
\end{equation}
which is very similar to the macroscopic equation (\ref{eq:evoDGL}) with the first term the probability of climbing and the second that of trading.
The correction $\sigma$ can be written as
 \begin{equation}
 \sigma = \left( \frac{1}{N} \sum_i G(c_i) p_i\right) -  \langle G \rangle \epsilon
 \end{equation}
where $\langle G \rangle = \frac{1}{N} \sum_i G(c_i)$ ($=G(\bar{c})$ in our case) is the mean climbing probability.
The differentiation is
\begin{equation}
 \dot{\sigma} = \left( \frac{1}{N} \sum_i G(c_i) \dot{p}_i\right) -  \langle G \rangle \dot{\epsilon}
\end{equation}
and substituting (\ref{eq:evoPropI}) and (\ref{eq:evoDGLKorrT}) yields
\begin{equation}
 \dot{\sigma} = \langle G \rangle \epsilon + f(1-\epsilon) (\langle G^{2} \rangle -\langle G \rangle \langle G \rangle) + \sigma (f \langle G \rangle - \epsilon) - f \nu(t)
\end{equation}
where $\nu(t) = \left( \frac{1}{N} \sum_i G^2(c_i) p_i\right) -  \langle G^2 \rangle \epsilon$ is a new higher order covariance term between the second power of $G^2(c_i)$ and the state probability $p_i$.

More generally, let us further denote the mean over powers of $G$ as 
\begin{equation}
\langle G^k \rangle := \frac{1}{N} \sum_i G^k(c_i)
\end{equation}
 and write
 \begin{equation}
 \sigma_k := \left( \frac{1}{N} \sum_i G^k(c_i) p_i\right) -  \langle G^k \rangle \epsilon.
 \end{equation}
When we express the dynamical equations for the correction term and then for the higher moment that emerges at each step, we arrive at the following infinite system of equations:
\begin{equation}
\nonumber
\begin{array}{ll}
\dot{\epsilon} & = f (1-\epsilon) \langle G \rangle-f \sigma_1-2\epsilon^2\vspace{4pt}\\
\dot{\sigma_1} & = \langle G \rangle \epsilon + f (1-\epsilon) (\langle G^{2} \rangle -\langle G \rangle \langle G \rangle) -f \sigma_{2} - \epsilon \sigma_1 + f \langle G \rangle \sigma_1\vspace{4pt}\\
\dot{\sigma_2} & = \langle G \rangle \epsilon +f (1-\epsilon) (\langle G^{3} \rangle -\langle G^2 \rangle \langle G \rangle) -f \sigma_{3} - \epsilon \sigma_2 +f \langle G^2 \rangle \sigma_1
\vspace{0pt}\\
& \vdots  \vspace{0pt}\\
\dot{\sigma_k} & =\langle G \rangle \epsilon + f (1-\epsilon) (\langle G^{k+1} \rangle -\langle G^k \rangle \langle G \rangle) -f \sigma_{k+1} - \epsilon \sigma_k +f \langle G^k \rangle \sigma_1
\end{array}
\end{equation}
Hence, we see that a relatively simple form of heterogeneity -- namely, a fixed heterogeneous cost threshold -- leads to a rather complicated system when we aim at a closed description in terms of macroscopic or average entities.
On the other hand, if $G(c_i)$ is strictly below one, these higher moments $\sigma_k$ tend to zero as $k$ increases which allows, under certain conditions, to close the system.
However, this goes beyond the scope of this paper and will be addressed elsewhere.

\subsection{Estimation of Correction Term From Simulations}

Another way to illustrate the usefulness of the correction term derived for the heterogeneous model is to estimate $\sigma$ from simulations.
Therefore, we run the ABM with a certain strategy distribution and compute $\sigma(t) = cov(s^t_i,G(c_i))$ for each step.
We then compute the time average over 2000 simulation steps which we may denote by $\bar{\sigma}$ and replace the correction terms in (\ref{eq:corrMC}) and (\ref{eq:evoDGLKorrT}) by this average heterogeneity term.
The respective transition probability for the Markovian description then becomes 
\begin{equation}
P(e+1 | e) = f \left[\frac{(N-e)}{N} G(c) - \bar{\sigma}\right]
\end{equation}
and we can compute the corrected stationary vector on that basis.
Likewise from (\ref{eq:evoDGLKorrT}) we obtain for the fixed point solution
\begin{equation}
\epsilon^*(c^*) = \frac{f G(c^*)}{4} \left[ \sqrt{1 + \frac{8}{f G(c^*)} - \frac{8 \bar{\sigma}}{f G(c^*)^2}} - 1 \right].
\label{eq:fixKorr}
\end{equation}

These terms are now confronted with simulations initialized according to different distributions of individual strategies.
Namely, we consider four cases:
\begin{enumerate}
\item strategies uniformly distributed within $[c_{min},c_{max}]$, 
\item there are two different strategies distributed at equal proportion over the population
\item the probability of a strategy $c_i$ decreases linearly from $c_{min}$ to $c_{max}$ and reaches zero at $c_{max}$
\item probability of a strategy $c_i$ decreases according to a $\Gamma$-distribution with shape $1$ and scale $1/5$
\end{enumerate}
Notice that the first two cases are chosen such that the mean climbing probability is $\langle G \rangle = 0.5$ whereas lower thresholds and therefore a lower average climbing probability are implemented with the latter two.

\begin{figure}[h]
\centering
\includegraphics[width=0.44\linewidth]{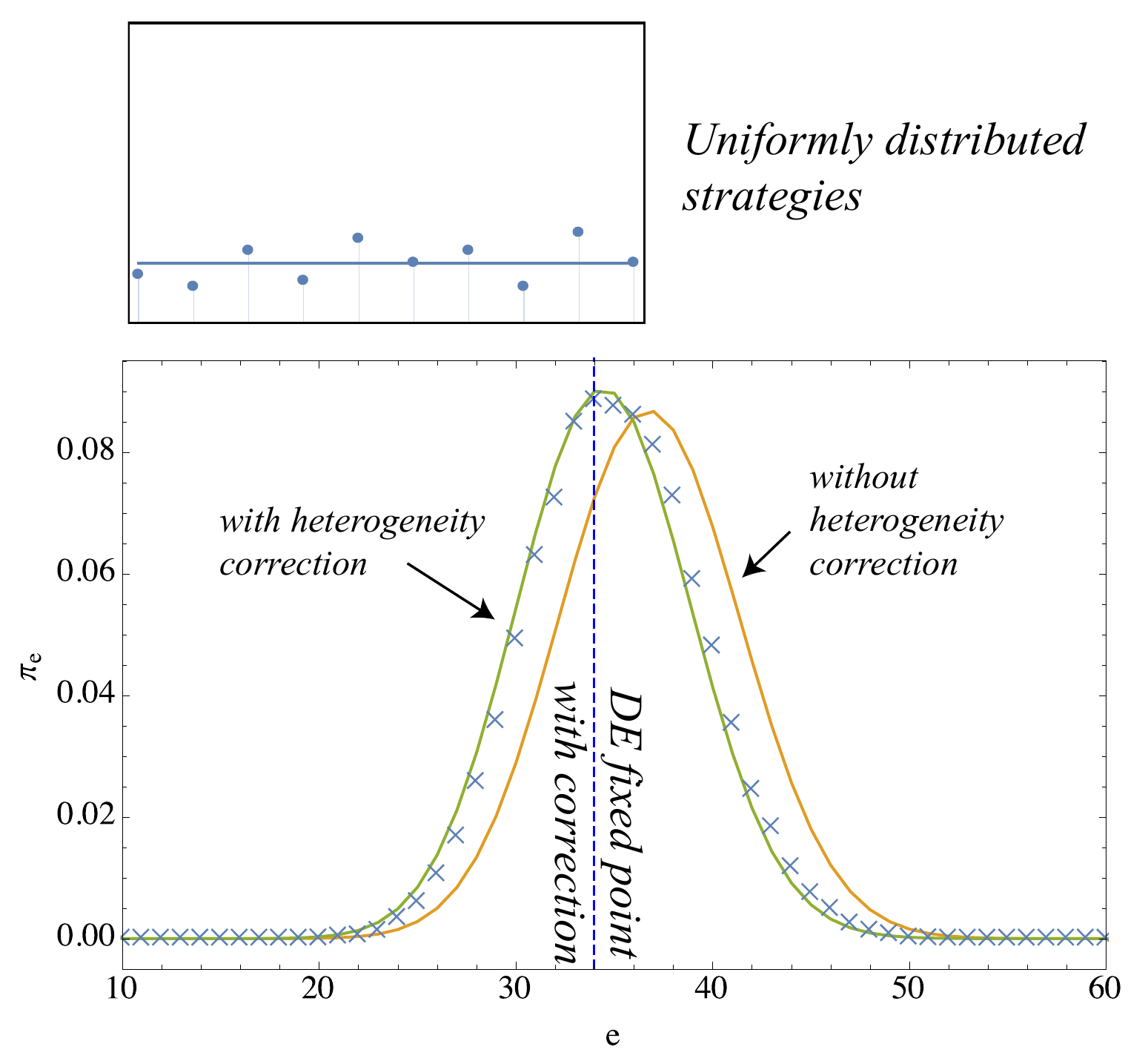} 
\includegraphics[width=0.44\linewidth]{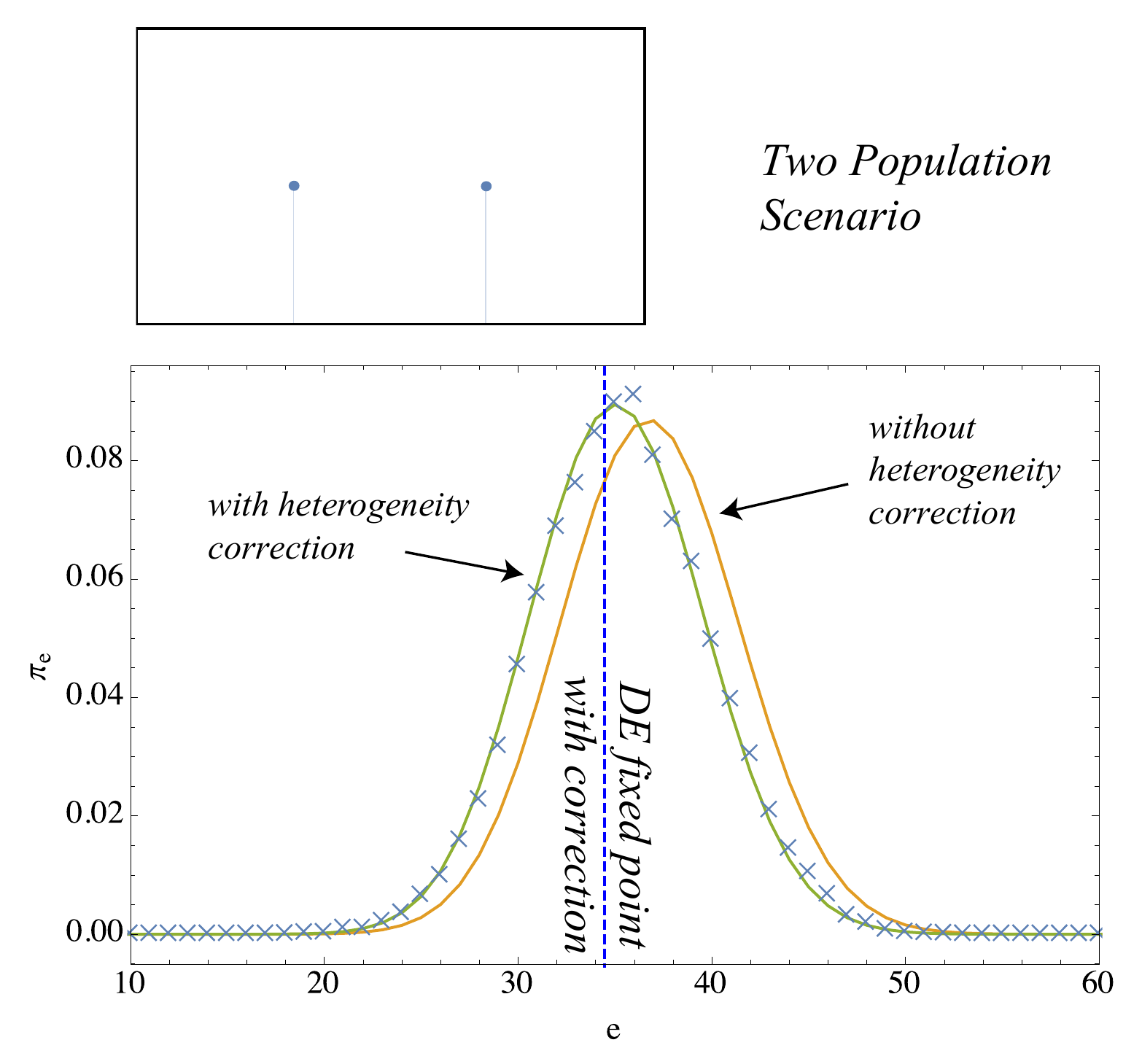} 
\caption{Stationary distribution for $N = 100$, $f = 0.8$, $c_{min}=0.3, c_{max}=0.5$ and the comparison to the frequencies observed in a series of 10 realization (10000 steps) of the simulation model.}
\label{fig:pi100Korr01}
\end{figure}

\begin{figure}[h]
\centering
\includegraphics[width=0.44\linewidth]{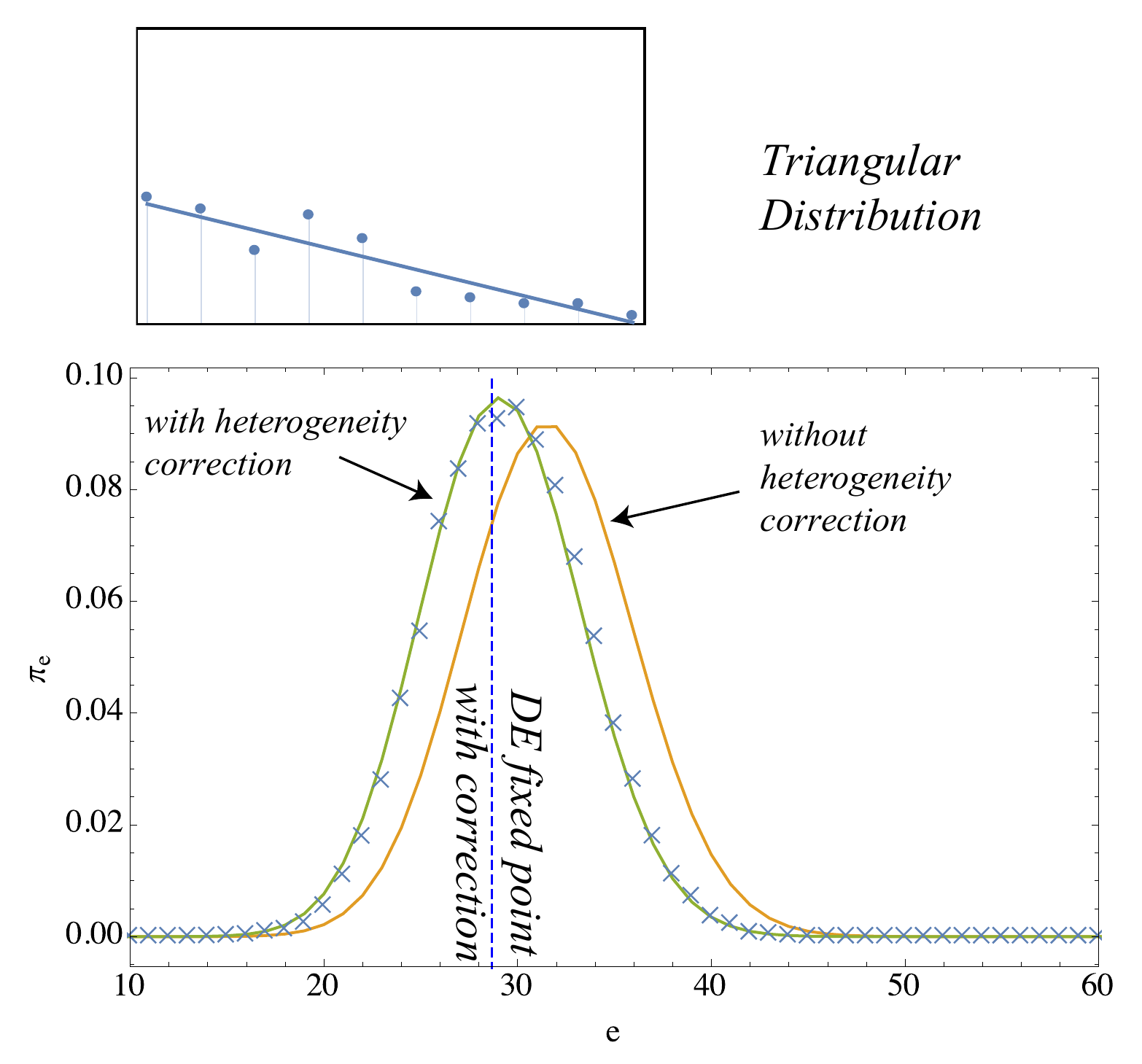} 
\includegraphics[width=0.44\linewidth]{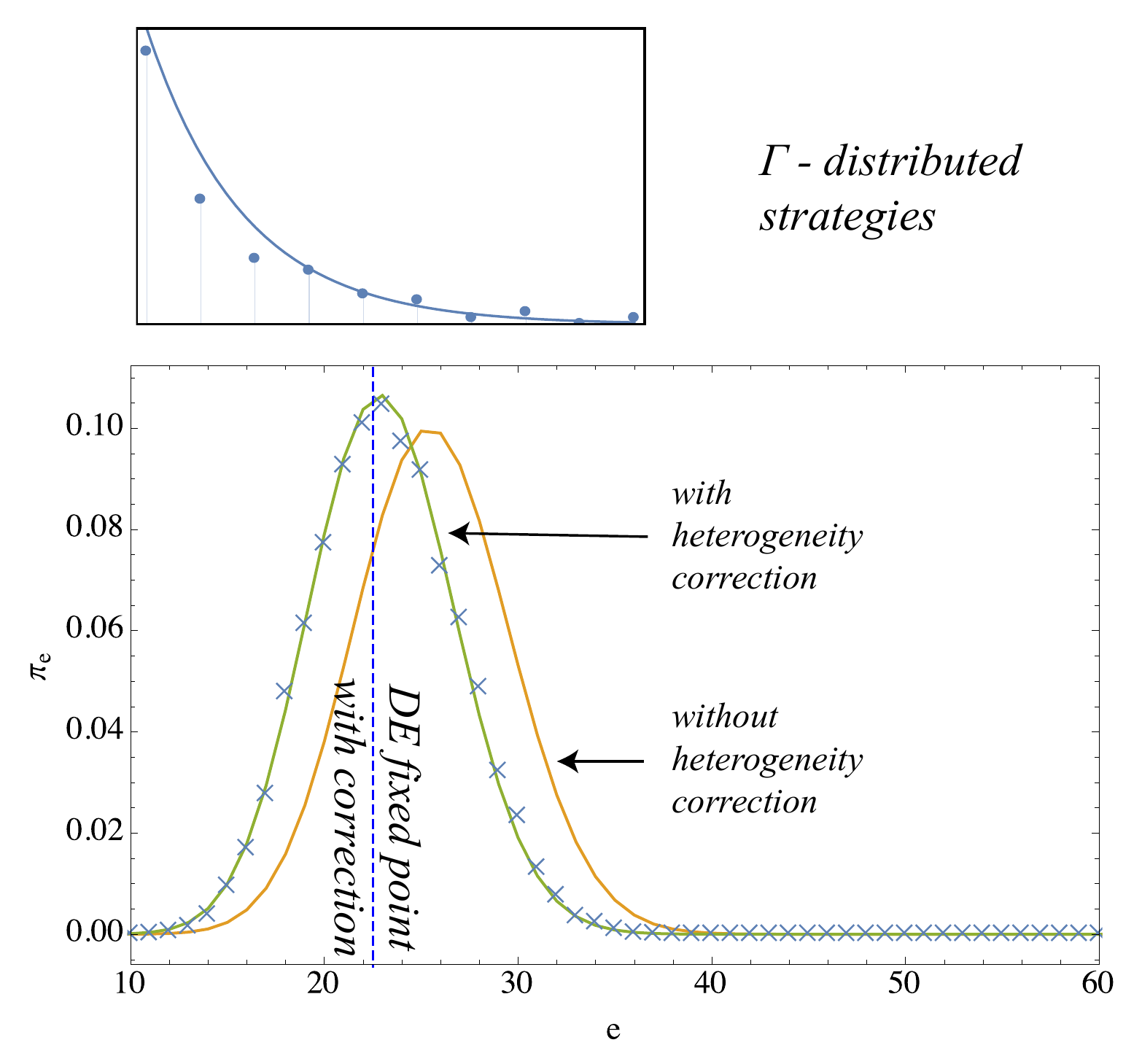} 
\caption{Stationary distribution for $N = 100$, $f = 0.8$, $c_{min}=0.3, c_{max}=0.5$ and the comparison to the frequencies observed in a series of 10 realization (10000 steps) of the simulation model.}
\label{fig:pi100Korr02}
\end{figure}

The results of applying the heterogeneity correction are shown in Figs. \ref{fig:pi100Korr01} and \ref{fig:pi100Korr02}.
For the four scenarios, the fixed point with correction (\ref{eq:fixKorr}) is shown by the vertical line along with the two stationary distributions of the MCs with and without correction.
The crosses in the plot correspond to simulation results measured on 10 simulation runs à 10000 steps.
In general, while significant deviations from the simulation results are visible for the descriptions without correction (homogeneous case), the stationary statistics of the model are well--matched after the covariance correction is applied.
This shows that very effective macroscopic formulations of heterogeneous agent systems may be possible by including correction terms that efficiently condense the actual micro-level heterogeneity in the system.

\section{Adaptive Strategies and Learning}
\label{sec:learning}

Having gained understanding about how to deal with heterogeneous strategies in the finite-size Coconut Model, we shall now turn to an adaptive mechanism by which the strategies are endogenously set by the agents.
As in Section \ref{sec:homogeneous}, we follow in this implementation the conception of Diamond \cite{Diamond1982} as closely as possible.
That is, firstly, the threshold $c_i$ has to trade off the cost of climbing against the expected future gain of earning a coconut from it.
In other words, agents have to compare the value (or expected performance if you wish) of having a coconut $V^t_i(1)$ with the value $V^t_i(0)$ of staying without a nut.
If the difference between the expected gain from harvesting at time $t$ and that of not harvesting ($V^t_i(1) - V^t_i(0)$) is larger than the cost of the tree $c_{tree}$, agents can expect a positive reward from harvesting a nut now.
Therefore, in accordance to \cite{Diamond1982}, it is reasonable to set $c_i^t = V^t_i(1) - V^t_i(0)$.

Now, how do agents arrive at reliable estimates of $V^t_i(1)$ and $V^t_i(0)$?
We propose that they do so by a simple temporal difference (TD) learning scheme that has been designed to solve dynamic programming problems as posed in the original model.
Notice that for single-agent Markov decision processes temporal difference schemes are proven to converge to the optimal value functions \cite{Sutton1998}.
In the Coconut Model with agents updated sequentially it is reasonable to hypothesize that we arrive at accurate estimates of $V^t_i(1)$ and $V^t_i(0)$ as well.
But notice also that the decision problem as posed in \cite{Diamond1982} is not the only possibility to formulate the problem.
Namely the model assumes that agents condition their action only on their own current state neglecting previous trends and information about other agents the consideration of which might lead to a richer set of solutions.
In our case agents do not learn the $\epsilon$ dependence explicitly, which means that the agents will only learn optimal stationary strategies.
The consideration of more complex (and possibly heterogeneous) information sets points certainly to interesting extensions of the model.
However, we think that it is useful to first understand the basic model and relate it to the available theoretical results as this will also be needed to understand additional contributions by model extensions.

\subsection{Learning the Value Functions by Temporal Differences}

The learning algorithm we propose is a very simple value TD scheme.
Agents use their own reward signal $r_i^t$ to update the values of $V^t_i(s_i^t=1)$ and $V^t_i(s_i^t = 0)$ independently from what other agents are doing.
In each iteration agents compute the TD error by comparing their current reward plus the discounted expected future gains to their current value estimate
\begin{equation}
\delta_i^{t+1} = \underbrace{r_i^{t+1}}_{\text{\normalsize reward}} +  \underbrace{e^{-\gamma N^{-1}}  \left[ s_i^{t+1} V_i^t(1) + (1-s_i^{t+1}) V_i^t(0)  \right]}_{\text{\normalsize estimated discounted future value}} - \underbrace{\left[ s_i^t V_i^t(1) + (1-s_i^t) V_i^t(0) \right]}_{\text{\normalsize current estimate}}
\label{eq:TDerror}
\end{equation}
which becomes
\begin{equation}
\begin{array}{ccl}
(0 \rightarrow 0): & \delta_i^{t+1} = & e^{-\gamma N^{-1}} V_i^t(0)  - V_i^t(0) 
\\
(0 \rightarrow 1): &\delta_i^{t+1} = & -c_{tree} + e^{-\gamma N^{-1}} V_i^1t(1) - V_i^t(0)
\\
(1 \rightarrow 0): &\delta_i^{t+1} = & y + e^{-\gamma N^{-1}}  V_i^t(0) - V_i^t(1)
\\
(1 \rightarrow 1): &\delta_i^{t+1} = & e^{-\gamma N^{-1}}  V_i^t(1) - V_i^t(1)
\end{array}
\end{equation}
for the different possible transitions $(s_i^t \rightarrow s_i^{t+1})$ of the agents.
Notice that the discount factor $\gamma$ as defined for the time continuous DE system is rescaled as $\gamma_r = e^{-\gamma N^{-1}}$ for the discrete-time setting and in order to account for the finite simulation with asynchronous update in which only one (out of $N$) agents is updated in each time step ($N^{-1}$).
The iterative update of the value functions is then given by 
\begin{equation}
\begin{array}{cl}
V^{t+1}_i(1) & = V^t_i(1) + \alpha \delta_i^{t+1} s_i^t
\\
V^{t+1}_i(0) & = V^t_i(0) + \alpha \delta_i^t (1-s_i^t)
\end{array}
\end{equation}
such that $V_i(1)$ ($V_i(0)$) is updated only if agent $i$ has been in state $1$ ($0$) in the preceding time step.

The idea behind this scheme and TD learning more generally is that the error between subsequent estimates of the values is reduced as learning proceeds which implies convergence to the true values.
The form in which we implement it here is probably the most simple one which does not involve update propagation using eligibility traces usually integrated to speed up the learning process \citep{Sutton1998}.
In other words, agents update only the value associated with their current state $s_i^t$.
While simplifying the mathematical description (the evolution depends only on the current state) we think this is also plausible as an agent decision heuristic.

All in all the model implementation is\vspace{6pt}\\
\begin{tcolorbox}[width=\textwidth,colback={aliceblue}]    
\begin{itemize}
\item[(0)] 
Initialization: set initial values $V^0_i(1),V^0_i(0)$ and states $s^0_i$ according to the desired initial distribution. Set initial strategies $c_i^0 = V^0_i(1)-V^0_i(0)$.
\item[(1)] Iteration loop I (search and trade):
	\begin{itemize}
	 \setlength\itemsep{-1pt}
	\item[(a)] random choice of an agent $i$ with probability $\omega(i) = 1/N$
	\item[(b)] \verb|if| $s_i = 0$ climb a coco tree with probability $f G(c_i)$ and harvest a nut, i.e., $s_i^{t+1} = 1$
	\item[(c)] \verb|else| trade (consume) with probability $\epsilon^t$ such that $s_i^{t+1} = 0$
	\end{itemize}
\item[(2)] Iteration loop II (learning): 
	\begin{itemize}
	 \setlength\itemsep{-1pt}
	\item[(a)] 
	compute TD error $\delta_i^{t+1}$ for all agents with reward signal $r_j = 0, \forall j \neq i$ and $r_i$ depending on the action of $i$ in part (1)
	\item[(b)]
	update relevant value function by $V^{t+1}_{i}(s_i^t) = V^t_i(s_i^t) + \alpha \delta_i^{t+1}$ for all agents
	\item[(c)] 
	update strategy by $c_i^{t+1} = V_i^{t+1}(1) - V_i^{t+1}(0)$
	\end{itemize}
\end{itemize}
\end{tcolorbox}

Notice that for trading we adopt the mechanism introduced as \verb|AM2| in Section \ref{sec:aligning}.
If not stated otherwise, the simulation experiments that follow are performed with the following parameters.
The interval from which the tree costs are drawn is given by $c_{max} = 0.5$ and $c_{min}=0.3$.
A strategy $c_i$ larger than $c_{max}$ hence means that the agent accepts any tree, $c_i<c_{min}$ that no tree is accepted at all.
The rate of tree encounter is $f=0.8$ and the utility of coconuts is $y = 0.6$.
We continue considering a relatively small system of 100 agent and the learning rate is $\alpha = 0.05$.
The parameter much of the analysis will be concentrated on is the discount rate $\gamma$ with small values indicating farsighted agents whereas larger values discount future observations more strongly.
The system is initialized (if not stated otherwise) with $\epsilon^0 = 0.5$, $V_i^0(1) = y$ and $V_i^0(0) = 0$ for all agents such that $c_i^0 = y > c_{max}$.

\subsection{Simulative Testing of Fixed Point Curves $\dot{c} = 0$ for Different $\gamma$}

\begin{figure}[h]
\centering
\includegraphics[width=0.55\textwidth]{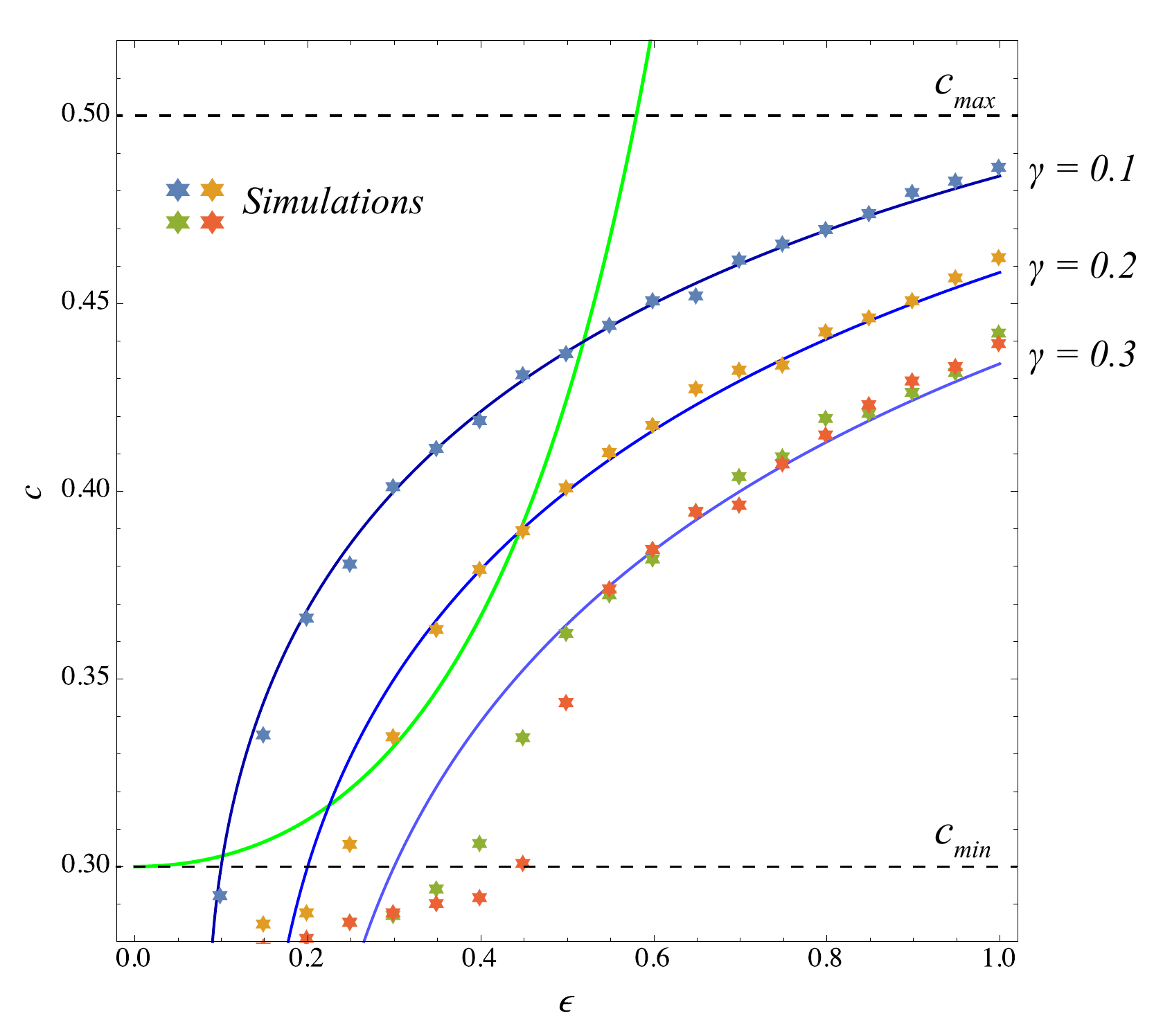}  
\caption{Fixed point curves for $\gamma = 0.1,0.2,0.3$ and comparison to the TD learning dynamics implemented in the agent model. }
\label{fig:LearningMatchTheory}
\end{figure}

The first part of this paper (exogonously fixed strategies) has shown that the ABM reproduces well the fixed point curve obtained for the coconut dynamics by setting $\dot{\epsilon} = 0$.
Here we want to find out whether the simulation with TD learning (adaptive strategies) matches with the fixed point behavior of the strategy dynamics of the original model obtained by setting $\dot{c} = 0$.
Fig. \ref{fig:LearningMatchTheory} shows the respective curves for three different $\gamma = 0.1,0.2,0.3$.
Notice that the last value $\gamma = 0.3$ is so large that the $\epsilon$ and $c$--curves do not intersect so that there is actually no fixed point solution.

In order to check these curves in the simulations we fix the expected probability of finding a trading partner by $b(\epsilon) = \epsilon_{fix}$.
Independent of the actual level of coconuts in the population, an agent finds a trading partner with that probability, consumes the coconut and derives a reward of $y$.
For Fig. \ref{fig:LearningMatchTheory}, for each $\epsilon_{fix} = 0.0, 0.05,0.1,\ldots,1.0$ the ABM is run a single time for 200000 steps and the last system configuration (namely, $c_i$ at the final step) is used in the computation of the mean strategy $c$ which is then plotted against $\epsilon_{fix}$.

The model generally matches with the theoretical behavior, especially when $\gamma$ is small (farsighted agents).
However, for $\gamma=0.2$ and $\gamma=0.3$ we observe noticeable differences between the simulations and the fixed point curve of the theoretical model.
Notice that the number of coconuts $\epsilon$ (which we fix for the trading step) actually also affects the probability with which an agent is chosen to climb and that the actual level of coconuts in the simulation is generally different from $\epsilon_{fix}$.
This might explain the deviations observed in Fig. \ref{fig:LearningMatchTheory}.
Setting up the experiment so that the level of coconuts is constant at $\epsilon_{fix}$, however, is not straightforward because an additional artificial state-switching mechanism would have to be included that has no counterpart in the actual model.

On the other hand, the results shown in Fig. \ref{fig:LearningMatchTheory} are actually a promising indication that agents which adapt according to TD learning align with the theoretical results in converging to the same (optimal) strategy for a given $\epsilon$.
The next logical step is now to compare the overall behavior of the ABM with learning to the theory.

\subsection{Overall Fixed Point Behavior with Learning}

\begin{figure}[h]
\centering
\includegraphics[width=1.0\textwidth]{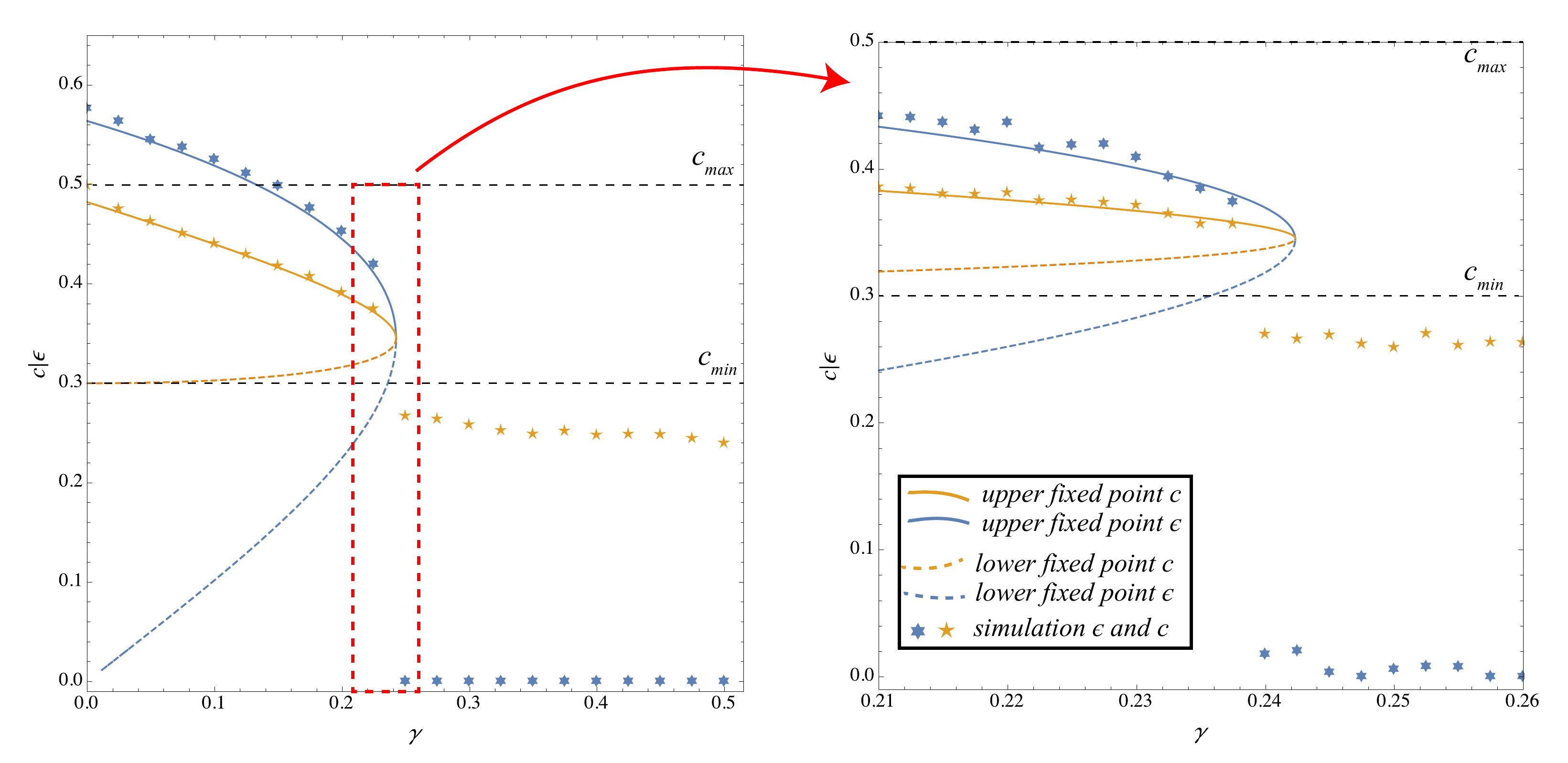}  
\caption{L.h.s.: The fixed point behavior of the DE system (\ref{eq:evoDGL}) - (\ref{eq:evoDGLStrat}) for $\gamma \in [0,0.5]$ is compared to single model realizations (200000 steps) for different $\gamma$. The agent model with  TD learning scheme converges closely to the theoretic fixed point values. R.h.s.: Close-up view $\gamma \in [0.21,0.26]$. In the experiments the model always reached the upper fixed point.}
\label{fig:FixedPointsGammaAI}
\end{figure}

For this purpose, we check the overall convergence behavior of the ABM as a function of $\gamma$ and compare it to the fixed point solution of \cite{Diamond1982}, see also \cite{Lux1992}.
There are two interesting questions here: 
\begin{enumerate}
\item 
what happens as we reach the bifurcation value $\gamma > \gamma^*$ at which the two fixed point curves $\dot{\epsilon} = 0$ and $\dot{c} = 0$ cease to intersect?
\item
in the parameter space where they intersect, which of the two solutions is actually realized by the ABM with TD learning?
\end{enumerate}
Both questions are answered with Fig. \ref{fig:FixedPointsGammaAI}.

First, if $\gamma$ becomes large, the ABM converges to the state in which agents do not climb any longer.
That is, $\epsilon^* = 0$ and $c^* < c_{min}$.
However, as the close-up view on the right-hand side shows, the bifurcation takes place at slightly lower values of $\gamma$.
This is probably related to the deviations observed in Fig. \ref{fig:LearningMatchTheory}.
In fact, further experiments revealed that the learning rate $\alpha$ governing the fluctuations of the value estimates plays a decisive role (the larger $\alpha$, the smaller the bifurcation point).
The larger $\alpha$ is, the more likely a perturbation takes place on the values of an agent ($i$) which takes $c_i < c_{min}$ meaning that agent $i$ does not climb any longer.
Besides this small deviation, however, Fig. \ref{fig:FixedPointsGammaAI} shows that on the whole the ABM reproduces the theoretical results with considerable accuracy.

Regarding the second question -- that is, equilibrium selection -- it seems that the only stable solution for the simulated dynamics is the upper fixed point, sometimes referred to as >>optimistic<< solution.
We will confirm this in the sequel by providing numerical arguments for the instability of the lower fixed point by a series of simulation experiments.

\subsection{Instability of the Lower Fixed Point}

The previous experiments indicate that the lower fixed point derived in the original system is generally unstable under learning dynamics. 
In this section we present some further results to confirm this observation by initializing the model at the lower fixed point.
We concentrate again on the parameterization used in the previous sections with $f=0.8, y=0.6, c_{min}=0.3, c_{max}=0.5$, climbing costs uniformly distributed in $[c_{min},c_{max}]$ and stick to a discount rate $\gamma = 0.1$.
As shown in Fig. \ref{fig:LearningMatchTheory}, the respective >>pessimistic<< equilibrium solution is given by $\epsilon^* \approx 0.102$ and $c^* \approx 0.303$ just slightly above $c_{min}$.
However, for the true fixed point initialization of the simulation model, we have to use the respective values at the fixed point $V^*(1), V^*(0)$ to initialize $V^0(1)$ and $V^0(0)$.
They can be obtained by solving the three-dimensional system from which Eq. (\ref{eq:evoDGLStrat}) has been derived \cite{Diamond1982}:
\begin{eqnarray}
\dot{\epsilon} = f (1-\epsilon) G(c) - \epsilon^2 \nonumber\\
\dot{V}(1) = \gamma V(1) + \epsilon(c- y) \nonumber\\
\dot{V}(0) = \gamma V(0) - f \int_{0}^{c} (c - c') dG(c').
\label{eq:CocoOriginalModel}
\end{eqnarray}
where $c = V(1)-V(0)$ as before.
With the parameters as given above, we obtain $V^*(1) \approx 0.303065$ and $V^*(0) \approx 0.000168$.

\begin{figure}[h]
\centering
\includegraphics[width=0.85\textwidth]{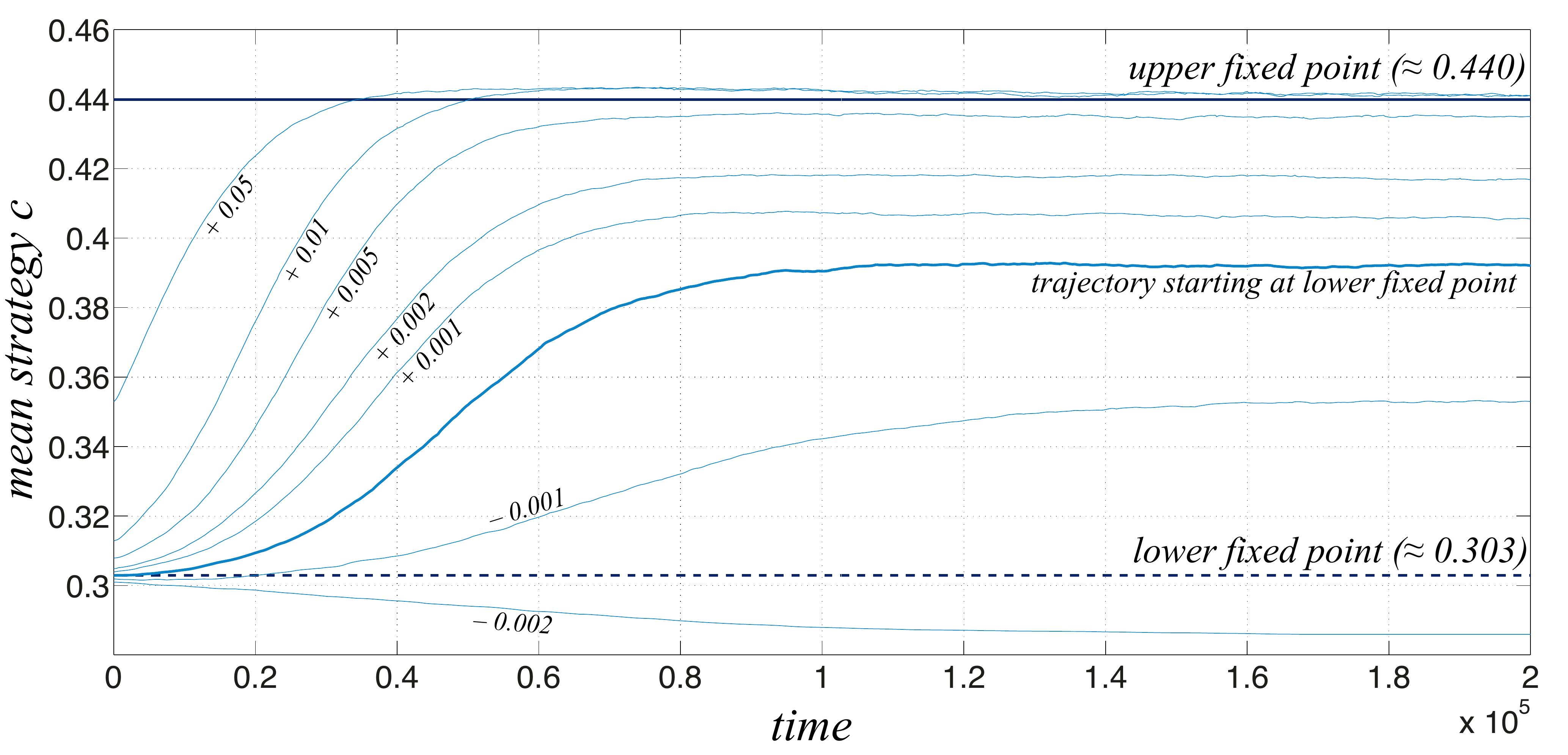}
\caption{200000 steps of the ABM with TD learning for various initial conditions close to the low fixed point (shown by the dashed dark line).}
\label{fig:stability01}
\end{figure}

Fig. \ref{fig:stability01} shows the evolution (200000 steps) of the ABM with TD learning for various initial conditions close to the low fixed point (shown by the dashed dark line).
There are 100 agents and the learning rate is $\alpha = 0.025$.
Each curve in the plot is an average over 5 simulation runs.
We first look at the bold curve corresponding to the trajectory starting at the lower fixed point (see indication in the figure).
It is clearly repelled from the low fixed point into the direction of the >>optimistic<< solution, however, it fails to reach the upper state.
The figure also shows trajectories that are initialized with slightly higher $V^0(1)$ leading to a slight increase of the initial strategy $c^0$.
The larger this initial deviation, the closer the trajectories converge to the expected >>optimistic<< strategy $c^* \approx 0.44$, but for a deviation smaller than $0.005$ the asymptotic behavior of the learning dynamics does not converge to this value.
Interestingly, also simulations initialized slightly below with an initial deviation of $V^0(1)=V^*(1)-0.001$ evolve into the direction of the upper solution, however, settling at a still smaller final strategy value. 

While this shows that the lower fixed point  (or at least a point very close to it) is repelling, this effect of convergence to almost arbitrary states in between the two expected fixed points seems somewhat surprising at the first glance.
One possible explanation is that TD learning schemes may converge to suboptimal solutions if agents to not sufficiently explore the space of possibilities.
In order to check if this is the reason for the unexpected convergence behavior when starting with the low initial values, we integrated a form of exploration by adding a small amount of noise to the strategies $c_i$ each time after the agents computed their new values.
This can be seen as if agents are not completely perfect in determining the values and the respective value difference $V(1)-V(0)$ or just as well that they believe to be able to form estimates only of a given finite precision.
In the simulations shown below, a random value uniformly distributed in between $-0.0015$ and $+0.0015$ has been added.

\begin{figure}[h]
\centering
\includegraphics[width=0.99\textwidth]{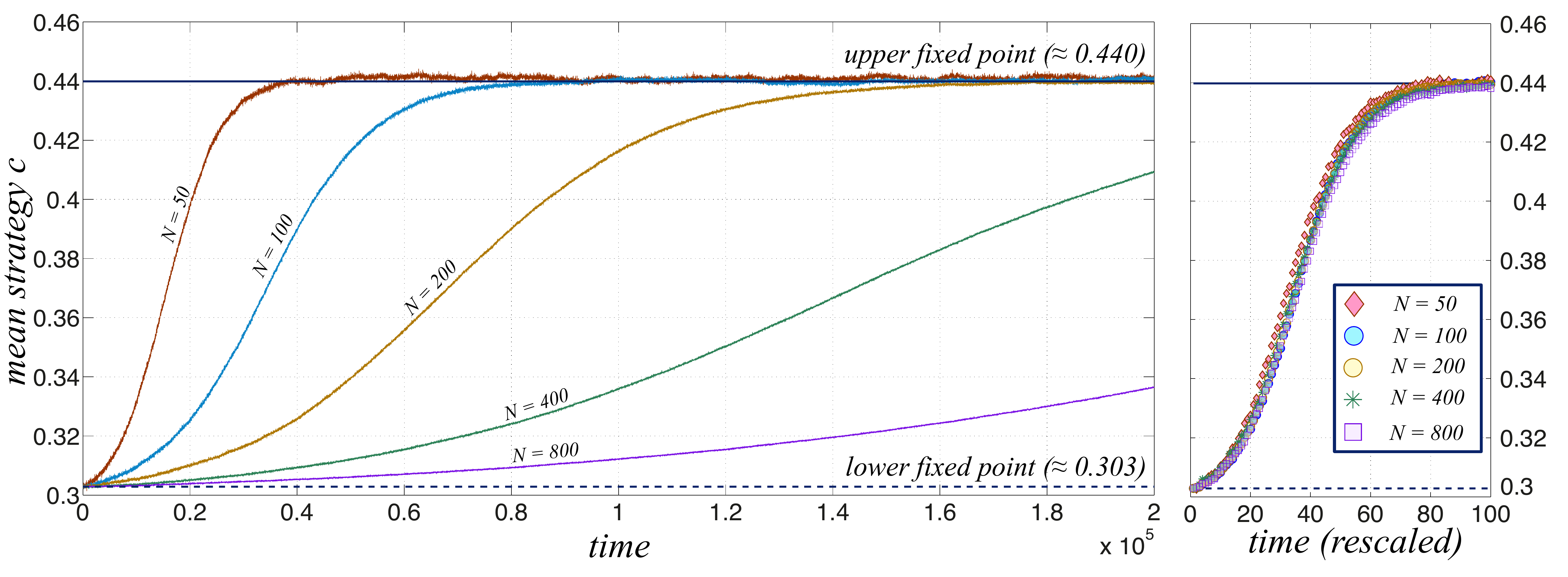}
\caption{R.h.s.: Time evolution of the system initialized at the low fixed point (shown by the dashed dark line) for different system sizes. L.h.s.: The same learning curve is obtained when rescaling time by the number of agents.}
\label{fig:stability02}
\end{figure}

The result of this >>exploration<< variant is shown in Fig. \ref{fig:stability02} which compares the time evolution of the model starting exactly in the theoretical low fixed point with respect to different system sizes.
In comparison with Fig. \ref{fig:stability01}, where only $N = 100$ has been considered and the initial deviation from the fixed point has been varied across realizations, we now observe an $S$--shaped curve that approaches the upper fixed point and stabilizes there with high accuracy.
This is observed for all $N$.

As the size of the system increases, the initial period in which the system stays close to the lower fixed point increases.
However, as shown on the right of Fig. \ref{fig:stability02}, the differences between the learning curves in systems of different size vanish when time is rescaled by the number of agents such that one time step accounts for $N$ individual updates.
This provides further evidence for the instability of the lower fixed point and shows that it is not merely a finite size effect but inherent in the agent system with procedural rationality based on TD learning.

\begin{figure}[h]
\centering
\includegraphics[width=0.49\textwidth]{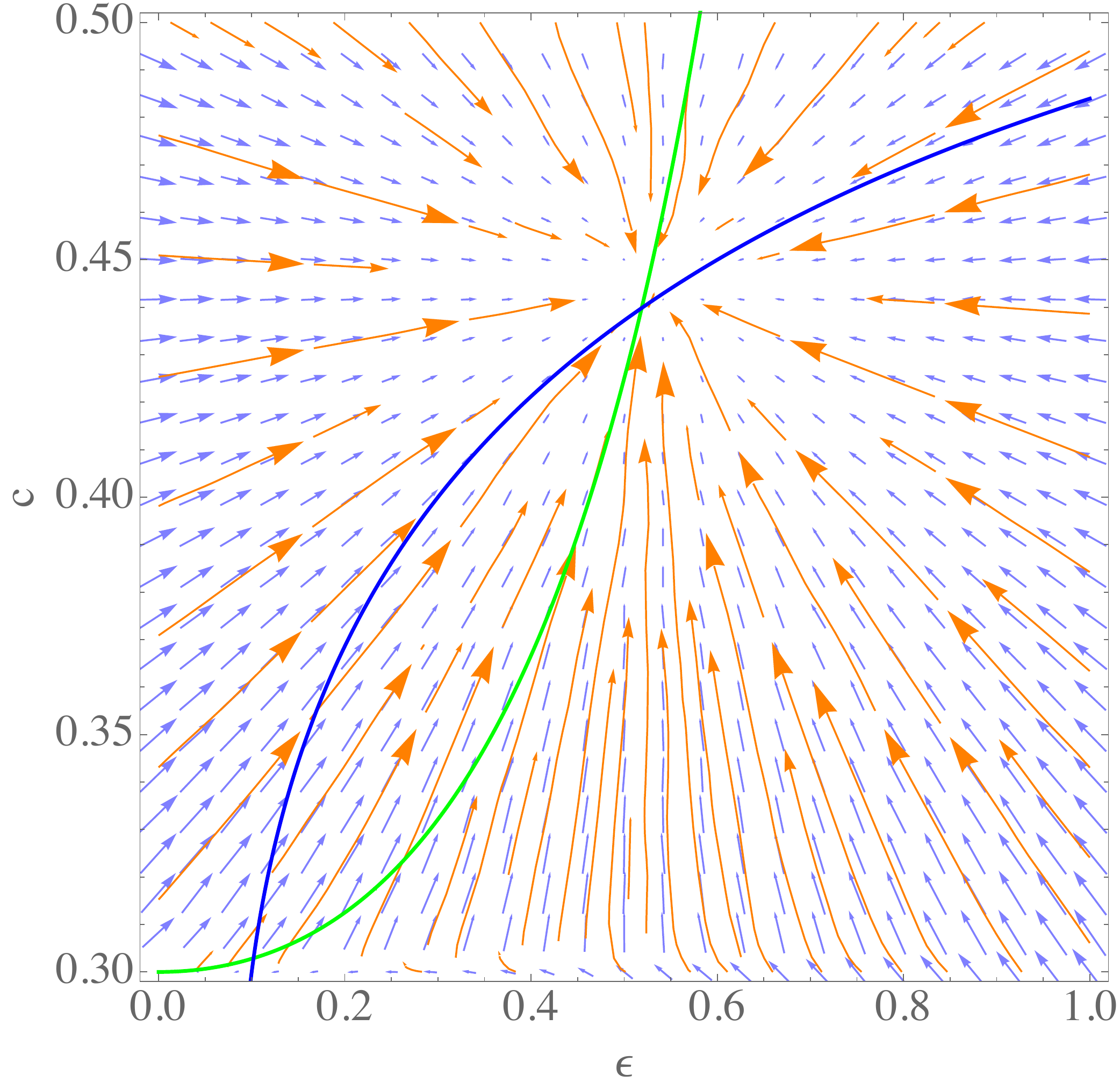}
\includegraphics[width=0.49\textwidth]{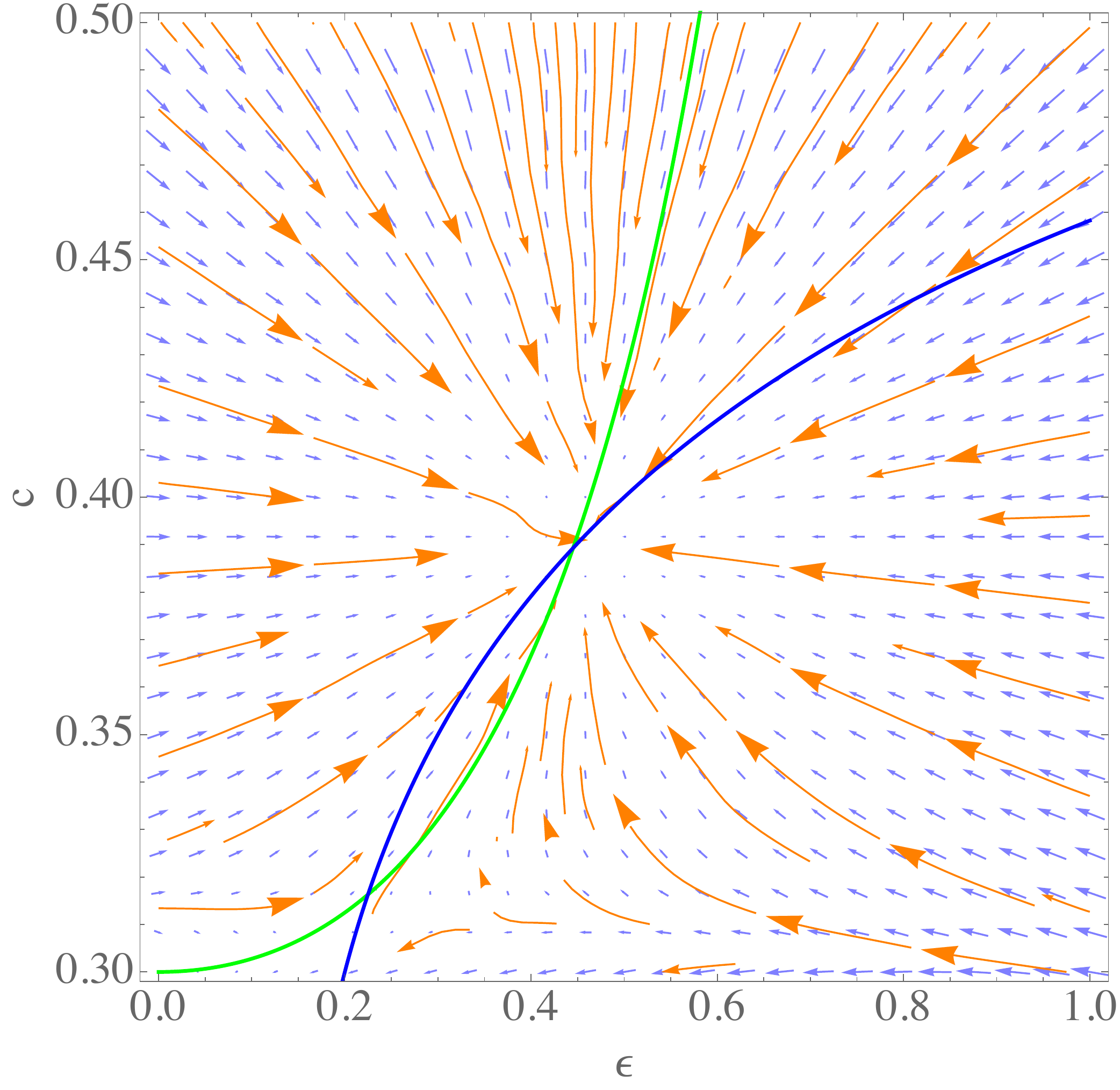}
\caption{Numerically constructed vector field for the dynamics of the agent model for $\gamma = 0.1$ (left) and $\gamma= 0.2$ (right). The fixed point curves of the DE system are also shown.}
\label{fig:stability03}
\end{figure}

To summarize the analysis performed in this section, we computationally constructed the phase diagrams for the dynamics with learning for a system of 100 agents.
That is, we perform a suite of systematic computations with samples of initial conditions in the plane spanned by $\epsilon^0 \in [0,1]$ and $c^0 \in [c_{min},c_{max}]$.
We compute $26 \times 26$ samples where $c^0$ is determined by letting $V^0(0) = 0$ and $V^0(1) = c^0$ set homogeneously for the entire population.
The initial level of coconuts is set randomly with $\epsilon^0 \in [0,1]$ such that the probability for each agent to have a coconut in the beginning is $\epsilon^0$.
For each initial combination we compute 10 simulations à 10000 steps and compare the initial point $(\epsilon^0,c^0)$ with the respective outcome after 10000 steps.
The result is shown in Fig \ref{fig:stability03} for a discount rate of $\gamma=0.1$ (l.h.s.) and $\gamma= 0.2$ (r.h.s.).

The vector field indicates convergence to a state close to the upper fixed point for most of the initial conditions.
For $\gamma = 0.1$ this is true even for very small initial strategies $c^0 < 0.303$.
However, we should notice that this point is very close to $c_{min}$ and that the sampling does not resolve the region around the low fixed point well enough.
For $\gamma = 0.2$ where the strategy value in the lower fixed point increases to $c^* \approx 0.316$ the dynamics around that point become visible.
In this case we observe that initial strategies below this value lead to convergence to $c < c_{min}$, that is to the situation in which agents do not climb any longer (and therefore $\epsilon = 0$).
However, if initially the level of coconuts is high enough the system is capable of reaching the stable upper solution because there is at least one instant of learning that having a nut is profitable ($V(1)$) for agents initially endowed with a nut.

\begin{figure}[h]
\centering
\includegraphics[width=0.59\textwidth]{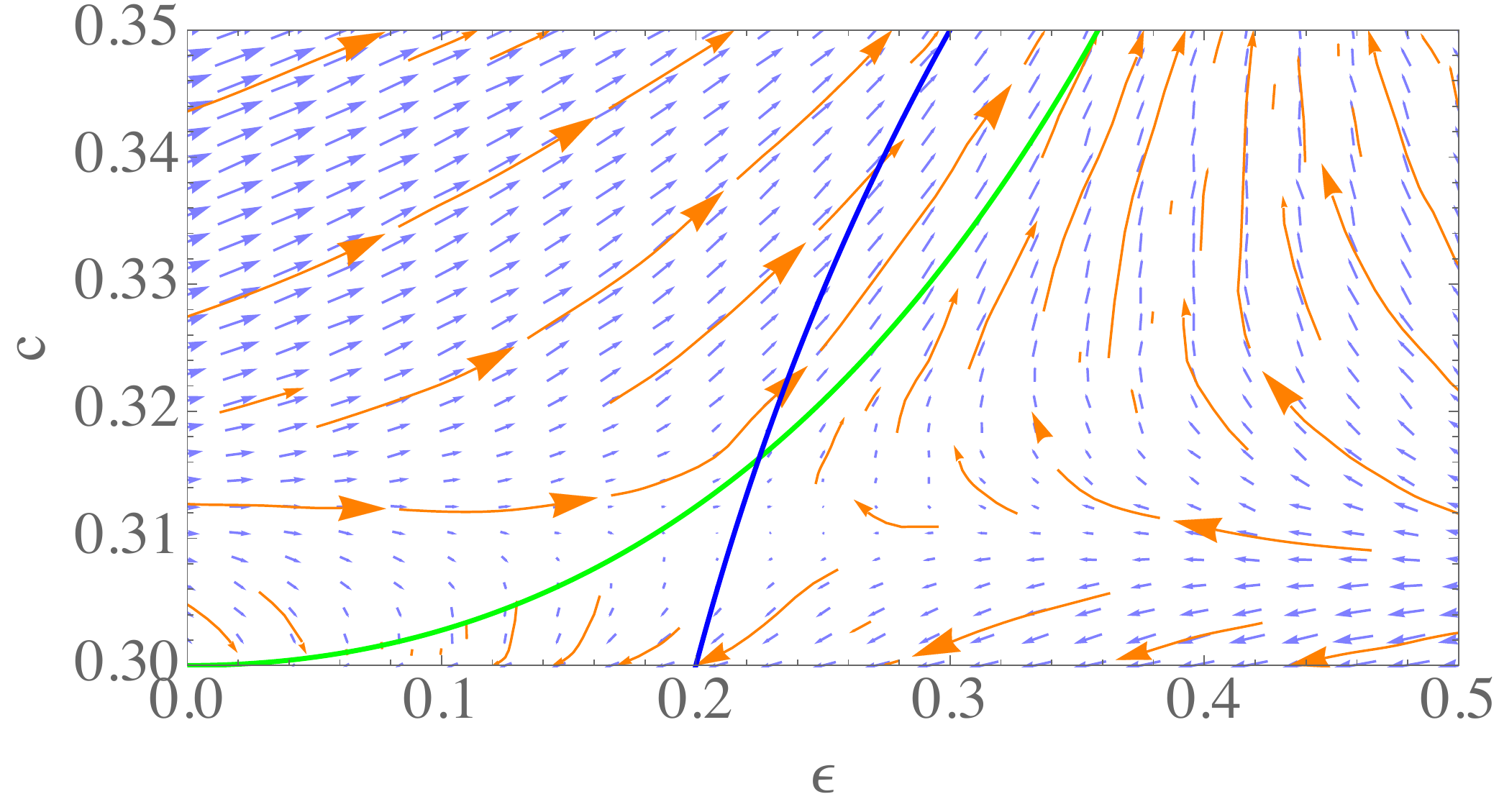}
\caption{Close-up view on the vector field. The lower fixed point acts as a saddle point under the learning dynamics.}
\label{fig:stability04}
\end{figure}

Finally, a close-up view on this region is provided in Fig. \ref{fig:stability04} for $\gamma = 0.2$.
It renders visible that the lower fixed point acts as a saddle under the learning dynamics.
As noticed earlier, the exact fixed point values $\epsilon^*$ and $c^*$ are slightly different for the DE system and the learning agents model which may be attributed to small differences in the models such as explicit exclusion of self-trading (see Section \ref{sec:homogeneous}) or the discrete learning rate $\alpha$ (this section).

\section{Outline and Conclusion}

This paper makes four contributions.
First, it develops a theory-aligned agent-based version of Diamond's coconut model \citep{Diamond1982}.
In the model agents have to make investment decisions to produce some good and have to find buyers for that good.
Step by step, we analyzed the effects of single ingredients in that model -- from homogeneous to heterogeneous to adaptive strategies -- and relate them to the qualitative results obtained from the original dynamical systems description.
We computationally verify that the overall behavior of the ABM with adaptive strategies aligns to a considerable accuracy with the results obtained in the original model.
The main outcome of this exercise is the availability of an abstract baseline model for search equilibrium which allows to analyze more realistic behavioral assumptions such as trade networks, heterogeneous information sets and different forms of bounded rationality but contains the idealized solution as a limiting case.

Secondly, this work provides insight on the effects of micro-level heterogeneity on the macroscopic dynamics and shows how heterogeneous agents can be taken into account in aggregate descriptions.
We derive a heterogeneity correction term that condenses the present heterogeneity in the system and show how this term should be coupled to the mean-field equation.
These mathematical arguments show that a full characterization of the system with heterogeneity leads to an infinite dimensional system of differential equations the analysis of which will be addressed in the future.
In this paper we have provided support for the suitability of the heterogeneity term by simulation experiments with four different strategy distributions.
We envision that the heterogeneity correction may be useful for other models such as opinion dynamics with heterogeneous agent susceptibilities as well.

The third contribution this paper makes, is the introduction of temporal difference (TD) learning as a way to address problems that involve inter-temporal optimization in an agent-based setting.
The coconut model serves this purpose so well because the strategy equation in the original paper is based on dynamic programming principles which are also at the root in this branch of reinforcement learning.
Due to this common foundation we arrive at an adaptive mechanism for endogenous strategy evolution that converges to one of the theoretical equilibria, but provides, in addition to that, means to understand how (and if) this equilibrium is reached from an out-of-equilibrium situation.
Such a characterization of the model dynamics is not possible in the original formulation.

Our fourth contribution relates to that in providing some new insight into equilibrium selection and stability of equilibria in the coconut model.
Under learning dynamics only the upper >>optimistic<< solution with a high coconut level (high productivity) is realized.
Furthermore, convergence to this equilibrium takes place for a great proportion of out-of-equilibrium states.
In fact, the phase diagrams presented at the end of the previous section show that in a system with farsighted agents ($\gamma = 0.1$) the market failure equilibrium (no production, no trade) is reached only if agents are exceedingly pessimistic.
If agents are less farsighted ($\gamma = 0.2$), this turning point increases slightly and makes market failure probable if the production level ($f G(c^t_i)$) is currently low for some reason.
However, we do not want to make general claims about the absence of cyclic equilibria in the artificial search and barter economy that the coconut model exemplifies.
It is possible -- even likely -- that a richer behavior is obtained when agents learn not only based on their own state but take into account information about the global state of the system, trends or the strategy of others.
This paper has been a necessary first step to address such question in the future.



\section{Model Availability}
 
 
All models described in this paper have been implemented and analyzed using Mathematica and Matlab. 
The Matlab version is made available at the OpenABM model archive \citep{CoconutModel2016} (See https://www.openabm.org/model/5045/version/1).



\end{document}